\documentstyle[amssymb]{JHEP3}

\makeatletter
\@addtoreset{equation}{section}
\makeatother

\def\dalemb#1#2{{\vbox{\hrule height .#2pt
        \hbox{\vrule width.#2pt height#1pt \kern#1pt
                \vrule width.#2pt}
        \hrule height.#2pt}}}

\def\tA{\widetilde A}
\def\tcA{{\widetilde{\cal A}}}

\def\hA{\hat{\cal A}}

\def\cA{{\cal A}}

\def\0{{\sst{(0)}}}
\def\1{{\sst{(1)}}}
\def\2{{\sst{(2)}}}
\def\3{{\sst{(3)}}}
\def\4{{\sst{(4)}}}
\def\5{{\sst{(5)}}}
\def\6{{\sst{(6)}}}
\def\7{{\sst{(7)}}}
\def\8{{\sst{(8)}}}

\def\tC{\widetilde C}

\def\tF{\widetilde F}
\def\tA{\widetilde A}

\def\R{\rlap{\rm I}\mkern3mu{\rm R}}
\def\G{{\cal G}}

\def\CS{{\cal S}}

\def\wtd{\widetilde}

\def\nn{\nonumber} \def\bd{\begin{document}} \def\ed{\end{document}}
\def\ds{\documentstyle} \let\fr=\frac \let\bl=\bigl \let\br=\bigr
\let\Br=\Bigr \let\Bl=\Bigl
\let\bm=\bibitem
\let\na=\nabla
\let\pa=\partial \let\ov=\overline
\newcommand{\be}{\begin{equation}}
\newcommand{\ee}{\end{equation}}
\def\ba{\begin{array}}
\def\ea{\end{array}}
\def\ft#1#2{{\textstyle{{\scriptstyle #1}\over {\scriptstyle #2}}}}
\def\fft#1#2{{#1 \over #2}}
\def\del{\partial}
\def\sst#1{{\scriptscriptstyle #1}}
\def\oneone{\rlap 1\mkern4mu{\rm l}}
\def\ie{{\it i.e.\ }}
\def\via{{\it via}}
\def\semi{{\ltimes}}
\def\v{{\cal V}}
\def\str{{\rm str}}
\def\Dm{{{D_{\sst{max}}}}}

\newcommand{\ho}[1]{$\, ^{#1}$}
\newcommand{\hoch}[1]{$\, ^{#1}$}
\newcommand{\bea}{\begin{eqnarray}}
\newcommand{\eea}{\end{eqnarray}}
\newcommand{\ra}{\rightarrow}
\newcommand{\lra}{\longrightarrow}
\newcommand{\Lra}{\Leftrightarrow}
\newcommand{\bp}{\tilde \beta^\prime}
\newcommand{\tr}{{\rm tr} }
\newcommand{\Tr}{{\rm Tr} }

\newcommand{\CP}{{\mathbb P}}

\newcommand{\NP}{Nucl. Phys. }
\newcommand{\ens}{\it Laboratoire de Physique Th\'eorique de l'\'Ecole
Normale Sup\'erieure\\
24 Rue Lhomond - 75231 Paris CEDEX 05}

\title{Borcherds symmetries in M-theory
\\ {\normalsize Dedicated to Pr. S. Hawking on his 60th birthday.
}}

\author{Pierre Henry-Labord\`ere, Bernard Julia and Louis Paulot\\
Laboratoire de Physique Th{\'e}orique
de l'Ecole Normale Sup{\'e}rieure\thanks{ UMR 8549 du Centre National
de la Recherche Scientifique et de l'\'Ecole Normale Sup\'erieure}\\ 
24  rue Lhomond, 75231 Paris Cedex 05, France\\
\email{phenry@lpt.ens.fr}\\
\email{bernard.julia@lpt.ens.fr}\\ 
\email{paulot@lpt.ens.fr}}

\abstract{
It is well known but rather mysterious that root spaces of the $E_k$ Lie groups
appear in the second integral cohomology of regular, complex, compact, del 
Pezzo surfaces.
The corresponding groups act on the scalar fields (0-forms) of toroidal 
compactifications of M 
theory. Their Borel subgroups are actually subgroups of supergroups of finite 
dimension over the Grassmann algebra of differential forms on spacetime that 
have been shown to preserve the self-duality equation obeyed by all
bosonic form-fields of the theory. We show here that the corresponding duality 
superalgebras are nothing but Borcherds 
superalgebras truncated by the above choice of Grassmann coefficients. 
The full  
Borcherds' root lattices are the second integral cohomology of the del 
Pezzo surfaces. Our choice of simple roots uses the anti-canonical form and 
its known orthogonal complement.

Another result is the determination of del Pezzo surfaces associated to
other string and field theory models. 
 Dimensional reduction on $T^k$ corresponds to blow-up of $k$ points in 
general position with respect to each other. 
All theories of the Magic triangle that reduce to the $E_n$ sigma model 
in three dimensions correspond to singular del Pezzo surfaces with
$A_{8-n}$ (normal) singularity at a point.
The case of type I and heterotic theories if one drops their gauge sector
corresponds to non-normal (singular along a curve) del Pezzo's.  
We comment on  previous encounters with Borcherds algebras 
at the end of the paper.
}

\preprint{LPT-ENS-02/20\\hep-th/0203070}

\begin{document}
\addtocontents{toc}{\protect\setcounter{tocdepth}{2}}

\section{Introduction}
U-duality groups are discrete, apparently arithmetic, groups of symmetries of 
quantum string models
in various compactifications. They act on the scalar fields which are 
themselves coupled to instantonic sources. 
In the classical supergravity (low energy) limit the symmetry group $G$ is the
real form of this discrete group $G_{\mathbb Z}$.
More precisely in the supergravity limit the moduli space of vacua for the 
scalar fields is the symmetric space 
$G/KG\equiv S$ where $KG$ is the maximal compact 
subgroup of $G$ and $S$ the solvable subgroup given by the Iwasawa decomposition
theorem. In the triangular gauge one is left
with a solvable algebra $S$ of symmetries. This algebra has been extended to a 
superalgebra in 1998 to include the other p-form fields which are coupled to 
(p-1)-branes. Recently Iqbal, Neitzke and Vafa \cite{inv}
have shown that 1/2 BPS brane 
types of M-theory compactified on (rectangular) tori almost exactly correspond 
to spherical (\ie genus 0) generators in the second cohomology of some 
associated del Pezzo surfaces.  The integral cohomology contains in the 
classical 
fashion the root lattice of the U-duality  algebra $G$. 

It turns out that the full 
cohomology  of these surfaces  spans the root lattice of a Borcherds 
superalgebra. Actually it is a Borcherds algebra for 10d IIB theory, 
specifically it is the rank 2 toy algebra studied by R. Slansky in \cite{sla}; 
the corresponding 
Cartan matrix is precisely given by the opposite of the intersection form
on the surface. More generally in the presence of fermionic simple roots of
length squared equal to
 one it turns out one must consider them as  isotropic roots
in order to fit into the Borcherds framework keeping the rest of the 
intersection form unchanged. The main effects of this choice are to prevent 
twice the odd roots to be roots and  to preserve the symmetry under the 
(real roots') Weyl group.
The degree truncation of \cite{inv} corresponds  to the possible values
of the degrees of differential forms that appear in the finite 
dimensional (super)algebra of symmetries of the supergravity approach 
\cite{cjlp2}. One Cartan generator can be eliminated as well: it does not
 couple
to any propagating field potential and corresponds to the anticanonical class
(see definition below).
 
Section 2 is devoted to mathematical prerequisites in a condensed but hopefully 
useful arrangement. In sections 3 and 4 we analyze M theory as well as
its toroidal compactifications  from the point of view 
of the (dual) del Pezzo (projective) surfaces.
Serre duality implements on the projective
surfaces the Hodge self-duality in spacetime, it is not a symmetry of the full
Borcherds algebra but only of the physical equations of motion.
Section 5 is the extension to type I or heterotic theories and this requires 
nonnormal (ie singular along a curve) surfaces. This may require the 
introduction of a normalizing surface as an auxiliary space, but we shall
work on the singular variety as much as possible.  
Their Borcherds algebras are constructed in two ways: 
firstly following  the regular case, namely starting  from the 
known U-duality algebra and adding the p-form fields, secondly by using 
the projection from type II to type I as the fixed point set of an involution  
which works for the U-duality algebras \cite{Jcar} and also for the full 
Borcherds superalgebras. 
Finally we prove that the set of subtheories called the Magic triangle 
(with simply laced split duality groups) also admit dual surfaces 
but now with normal singularities. We explain the symmetry of the triangle 
with respect to the diagonal by using the correspondence between $A_n$ systems
of divisors and the toroidal chain of compactifications. The structure can in 
fact be generalised to many other triangles. We also discuss some alternative 
theories that arise in the systematic combination of regular and singular
contractions.
In the concluding section we recall other occurrences of Borcherds algebras and 
present a program for further work.

\bigskip
\section{Geometrical prerequisites}

 All the surfaces involved in Physics via this paper will be complex, algebraic 
and projective surfaces, \ie  projective varieties of complex dimension
two. An algebraic (projective) variety is a complex manifold which can
be described as the zero locus of some homogeneous polynomials in the $(n+1)$
coordinates of points of $\CP^n$ (for some $n$), in particular it is compact, 
but possibly singular. One may think of these surfaces as 
some kind of twistor spaces and then  of spacetimes as derived objects.
In fact they became relevant for string theory in \cite{inv} and we refer to 
that paper for introductory material and a partial dictionary of 
physical objects. The connections between $E_n$ root lattices, del Pezzo
surfaces and maximal supergravity on tori are of course much older.
For the sake of brevity we shall for the most part restrict our general 
considerations to regular surfaces  in this section, the details and subtleties 
of the general case are only alluded to here and left for later publications.

\subsection{Divisors and their classes}

On a compact algebraic variety X, a {\it Weil divisor} \cite{har,gri} 
is a finite formal linear
combination $\sum a_i V_i$ of irreducible analytic subvarieties of
complex codimension one $V_i$ with
integral coefficients $a_i$. If all $a_i$'s are positive, the divisor
is called effective. If $f$ is a meromorphic function on X, its zeroes
define codimension one subvarieties $Z_i$ (with zeroes of respective order 
$a_i$'s) and poles define 
subvarieties $P_j$ (of order $b_j$). One then associates to $f$ a divisor
$(f)=\sum a_i Z_i - \sum b_j P_j$. Such divisors are called {\it
principal}. Two divisors are {\it linearly equivalent} if their
difference is principal, and one can mod out the group of Weil
divisors  $Div(X)$ by principal (Weil) divisors to obtain the {\it divisor 
class group} of $X$: $Cl(X)$.

One can also define {\it Cartier divisors} as global sections of the
quotient (multiplicative) sheaf $\mathcal{M}^* / \mathcal{O}^*$, where
$\mathcal{M}^*$
is the sheaf of meromorphic functions on X not identically zero and
$\mathcal{O}^*$ the sheaf of nonzero holomorphic
functions on X. (Local sections of a sheaf are nothing but the defining 
functions or objects over an open set of the base.)   

Meromorphic functions trivially define Cartier divisors
which are still called {\it principal} and induce linear
equivalence. In fact, Cartier divisors on X are those Weil divisors
which are locally principal (with the intuitive meaning of locally). 
If X is smooth all Weil divisors are Cartier. So the notion of Cartier 
divisor becomes important in the singular case. 

Given a Cartier divisor D on X, one can construct a line bundle on X in the
following way. As  D is locally principal it can be 
represented by meromorphic functions $f_i$ on
an open cover $\{U_i\}$ of X, with $f_i/f_j$ holomorphic and
non-vanishing on $U_i \cap U_j$, one may then take $f_i/f_j$ as the transition
functions defining the associated line bundle on X. Linearly equivalent Cartier 
divisors give
isomorphic line bundles, and one gets a morphism from the Cartier divisor
class group ( $CaCl(X)$)
 into the group of isomorphism classes of line bundles on X (which
is called the {\it Picard group} Pic(X)). In fact this
is an isomorphism when X is projective as we have assumed here \cite{nak63}. 

\subsection{Intersection on algebraic surfaces}

On a {\it normal} surface (which means that the singular locus on X has 
codimension strictly larger than one) X, divisors are generated by
curves, and one can define an intersection number between divisors. 
If we restrict ourselves at first to smooth surfaces,
for smooth curves intersecting transversally the intersection number is simply 
the (nonnegative) number of intersection points and this
extends to other divisors. More precisely, there exists a unique
symmetric additive pairing $Div(X) \times Div(X) \ra {\mathbb Z}$
which depends only on the linear equivalence classes and reduces to
the number of intersection points for nonsingular curves meeting
transversally. Even for nonnormal projective surfaces the pairing 
is still defined and integer on Cartier divisors.
\cite{deb} (prop. 1.8).

If C and D are two curves on X and if one writes $c_1(Z)$ for the
first Chern class of the line bundle associated to a curve Z, this
pairing 
can be expressed as 
\be
C.D=\int_D c_1(C)=\int_C c_1(D)=\int_X c_1(C) \cup c_1(D) {\textrm .}
\ee

On a smooth compact rational surface, the intersection matrix is unimodular,
this 
reflects Poincar\'e duality. In the singular case the intersection matrix 
may have a Kernel and dividing out by the Kernel is called numerical 
equivalence. The nondegenerate part of the signature is recalled at the 
beginning of section 3. 

\subsection{Ample divisors, projective embeddings and degrees}

A line bundle $\mathcal L$ on an algebraic surface 
X is called {\it very ample} if, for some n, it has
$n+1$ linearly independent global
sections that can be used to define  an embedding of X in $\CP^n$ and 
$\mathcal L$ is then
the pull-back to X of the tautological bundle ${\mathcal O}(1)$ over
$\CP^n$. Moreover, any projective embedding of X is given in that way.
A line bundle is {\it ample} if it has a finite (positive) tensorial
power which is very ample. On a normal surface we shall say that a divisor is 
ample or very ample if its associated line bundle is.

Given an ample divisor H on a regular surface X, one defines the (H-)degree of 
any divisor C as $H.C$. This degree $d_H$ gives a morphism from the 
${\mathbb Z}$-module Pic(X) to ${\mathbb Z}$. If H is very ample its degree
$H.H$ is equal to the (algebraic) degree of the corresponding projective 
embedding. 
The {\it Nakai-Moishezon criterion} asserts that a divisor H is ample if and
only if $H^2(=H.H)>0$ and $H.C>0$ for any irreducible curve C.
We shall consider singular normal projective surfaces but define the degree
of Cartier divisors by requiring the 
corresponding ample divisor itself to be Cartier. 

\subsection{Canonical class $K_X$}

For a normal complex variety X of dimension $n$, the $n$-th tensorial power of
the cotangent bundle is a line bundle, and one can therefore
associate to it a divisor class $K_X$, the {\it canonical class}. Its
dual the tangent bundle is associated to the {\it anticanonical} class $-K_X$.

An important relation which holds for any nonsingular curve $C$ on a
smooth
algebraic projective surface X is known as the {\it Adjunction Formula}:
\be
(K_X+C).C=-2+2g(C)
\ee 
where $g(C)$ is the arithmetical genus of $C$ it coincides with the
geometrical 
(of Riemann surface theory) genus of $C$ for a regular irreducible
curve, 
sometimes we shall just call it the genus. Note that relation $(2.2)$
holds also for all normal singular surfaces considered in this work,
which have only {\it Du Val} singularities (this term will be
explained in section {\bf $6.2$}).

The {\it virtual genus} of any divisor $C$ (effective or not)
can be defined with this formula as
\be
g_v(C)=1+\frac{(K_X+C).C}{2} {\textrm .}
\ee
It obviously reduces to the geometrical genus for any nonsingular irreducible 
curve on a smooth surface.
One notes that for any rational (\ie of genus 0)  divisor $C$ 
its Serre dual: $-K_X-C$, is also rational.

\subsection{Blowing up points}

For ${\mathbb C}^2$, one defines the {\it blow up} of the origin as the 
surface X defined by 
$\{(z,l) \in {\mathbb C}^2 \times \CP^1 | \forall i,j \, \,  z_i l_j = z_j
l_i\}$. In other words, the origin is replaced with the $\CP^1$
of all directions of
lines passing through it. As this procedure is local, one can blow up
in the same way any smooth point P of an algebraic surface X. If Y is
the surface thus obtained, we get a projection morphism $\pi : Y \ra X$ such
that $\pi^{-1}(P) \simeq \CP^1$ and $Y \smallsetminus \pi^{-1}(P) \simeq
X \smallsetminus P$.

Denoting by  $E$ the divisor class of the {\it exceptional curve}
$\pi^{-1}(P)$, one can prove that $E$ has
self-intersection $E^2=-1$ and is perpendicular to the pull back of
any divisor of X not passing through P. We also have the relation
between the canonical classes
$K_Y=\pi^* K_X + E$ from which follows $K_Y^2=K_X^2-1$.

Conversely, if $E$ is a smooth curve isomorphic to $\CP^1$ and of
self-intersection $-1$, the {\it Castelnuovo-Enriques criterion} asserts
that Y can be blown down to a surface X such that $E$ is contracted to
a smooth point P and Y is precisely the blow up of X at the point P as
described above.

More generally if a curve $E$, still isomorphic to $\CP^1$, is of
self-intersection $-n$, it 
can be contracted to a point if and only if $n$ is positive. If $n$ is
larger than 1, we get a singular point. The relation between
canonical divisors  generalizes to $K_Y=\pi^* K_X + (2-n) E$. (In what
follows, we often omit $\pi^*$ when there is no ambiguity.)

\section{From del Pezzo surfaces to Borcherds superalgebras}

A {\it (generalized) del Pezzo surface} $X$ is by definition a connected
surface that is possibly singular but {\it Gorenstein} (\ie its 
anticanonical class $-K$ is a Cartier divisor) and is such that 
$-K$ is ample (hence X is projective).

We consider the vector space $Pic(X) \otimes_{\mathbb{Z}} \mathbb{R}$
over $\mathbb{R}$ of dimension $n$ with the induced bilinear form given by the
intersection matrix. According to the {\it Hodge Index Theorem}, the
signature of the bilinear form modulo its Kernel is 
$(1,n-1)$. $-K$  is a positive (or timelike) direction as 
it is ample by definition. 

Here comes our basic rule, we want to extend the structure of the instantons
in the orthogonal hyperplane to $K$ to other rational curves, this suggests the following procedure.
Let us assume that the set of {\it positive roots} contains the rational 
(\ie of vanishing virtual genus) divisor classes
of nonnegative degree  and also (except in the case of M-theory) the
anticanonical
class $-K$, which has also a positive degree by the
 del Pezzo defining property. 
We must now choose among the set of positive roots a  basis of
{\it simple roots} $\alpha_i$ such that any positive root can be written as a
linear combination of simple roots with positive, integral coefficients. In
fact, we observe that these $\alpha_i$'s generate the full Picard lattice of the
del Pezzo surface X.
$-K$ will turn out to have some 
multiplicity equal to the  rank of the root space minus one (this will be
related by Hodge 
duality to the fact that we introduce as many scalar fields 
as the number of generators of a codimension one Cartan subspace). 
In degree zero one is handling a usual Lie root lattice and
one can use the classical positive root decomposition. 
We note that all simple roots verify $\alpha_i^2 \geq -2$. For rational
divisors it follows from the adjunction formula and the positiveness of their
degrees. One sees that the same bound holds also for $-K$.
% This bound is in effect  a generalized simply-laced property.

We can now extract minus the {\it intersection matrix }  
$A_{i j}=-\alpha_{i}.\alpha_{j}$ and a 
${\mathbb Z}$ (resp. ${\mathbb Z}_2$)-graduation
$grad$ given by $grad(\alpha_i)=-K.\alpha_i \equiv d_{-K}(\alpha_i)$
 (resp. mod 2) from the cohomology.
By the adjunction formula the ${\mathbb Z}_2$-grading is precisely the
squared norm of the divisor mod 2. This ${\mathbb Z}_2$ parity of
the degree will correspond to 
the parity of the degree of the field potential in the supergravity theory.
In other words the fermionic character of the roots is dictated by the 
cohomology multiplicative structure.

However it turns out that whenever 
a fermionic divisor (root) of square $-1$ appears it should be 
viewed as an $SL(1|1)$ superroot rather than an $OSp(1|2)$ 
superroot, i.e. it should have zero Cartan(-Killing) norm. 
Furthermore the theory of Borcherds superalgebras is best developped for  the 
case of null simple fermionic roots. We must 
assume the generating fermionic simple root(s) 
(there are at most two of them in a diagram) to be isotropic but we keep 
its (their) intersection values with the other generators and the rest of the 
intersection form as given from cohomology. In other words, we put
$a_{ii}=0$ for fermionic roots which correspond to divisors of square
$-1$ but do not change the rest of the Cartan matrix. So the only
simple roots of positive norm are bosonic.

The corresponding modified matrix will be our
{\it Cartan matrix} $a_{ij}$.
This matrix $a_{ij}$ satisfies the following properties 
and thus defines a {\it Borcherds superalgebra} (or {\it Generalized Kac-Moody}
  superalgebra), without real fermionic simple roots\cite{bor, ray}.
\begin{eqnarray}
(i) & a_{ij} \leq 0 & {\mathrm if } \ i \neq j \\
(ii) & \frac{2a_{ij}}{a_{ii}} \in  {\mathbb Z} & {\mathrm if } \
a_{ii} > 0 
%\ ,\ {\mathrm for} \ grad(\alpha_i)=0
%\\
%(iii) & \frac{a_{ij}}{a_{ii}} \in {\mathbb Z} & 
%{\mathrm if } \ a_{ii}
%> 0 \ 
%, \ {\mathrm for} \ grad(\alpha_i)=1
\end{eqnarray}

We do not know of any conceptual proof of that fact but
as far as (ii) is concerned,  
it expresses the integrality of the corresponding
$a_{ij}$'s because $-\alpha_{i}^2 \leq 2$ here.
A similar correspondence between the Picard lattices of  $K3$ surfaces and
generalized Kac-Moody superalgebras without odd ``real'' (ie of strictly 
positive $a_{ii}$) simple roots has
been considered in \cite{nik}.
 
The Borcherds superalgebra associated to the matrix $a_{ij}$
has a Cartan subalgebra $H$, with basis $\{h_{\alpha_i}\}$, it is  by 
definition the Lie superalgebra $G$ generated by $H$ and by the 
elements $e_{\alpha_i}$, $f_{\alpha_i}$ satisfying here the following 
elementary relations and their consequences:

\begin{eqnarray}
(1) & [ e_{\alpha_i} ,f_{\alpha_j}  ] = \delta_{ij} h_{\alpha_i} \\
(2) & [ h_{\alpha_j} , e_{\alpha_i} ] = a_{ij} e_{\alpha_i}
\textrm{,} \,  [ h_{\alpha_j} ,
f_{\alpha_i} ] = -a_{ij} f_{\alpha_i} \\
(3) & [ h_{\alpha_i} , h_{\alpha_j} ] = 0 \\
(4) & {ad(e_{\alpha_i})}^{1-2\frac{a_{ij}}{a_{ii}}} e_{\alpha_j} = 0 
= (ad(f_{\alpha_i}))^{1-2\frac{a_{ij}}{a_{ii}}} f_{\alpha_j}
\ \textrm{if} \ a_{ii} > 0 \\
(5) & [ e_{\alpha_i} , e_{\alpha_j} ] = 0 = [ f_{\alpha_i} ,f_{\alpha_j}  ] 
\ \textrm{if} \ a_{ij} = 0 
\end{eqnarray}   
\\
In the following sections, we will compute the Borcherds superalgebra for each
del Pezzo surface $X$. Then, by truncating the set of positive superroots,
we will recover a superalgebra that preserves the classical supergravity 
equations of motion. The roots orthogonal to the canonical class are related 
to the usual duality groups even in the singular surface 
cases we shall encounter.
The divisors with vanishing virtual genus will correspond  to $1/2$
BPS states of these theories.
   
\section{Smooth del Pezzo's}
The smooth del Pezzo surfaces are
classified according to  what turns out to be the uncompactified spacetime 
dimension $d=K.K+2$ \cite{har,man}:
\begin{center}
\begin{tabular}{r l l}
(i) & if d=11 & $X \simeq {\mathbb P}^2$ \\ 
(ii) & if d=10 & $X \simeq \CP^1 
\times  \CP^1$ or $X \simeq {\mathbb B}_1$  \\
(iii) & if $3  \leq d = 11-k  \leq 9$ & $X \simeq
{\mathbb B}_k$ 
\end{tabular}
\end{center}
where ${\mathbb B}_k$ is the surface obtained by blowing up $k\le 8$ points
in general 
position on ${\mathbb P}^2$.

\subsection{M theory}
We will show that this theory corresponds to the regular 
surface ${\mathbb P}^2$.
First, we note that $Pic({\mathbb P}^2) \simeq {\mathbb Z} $ and we can take
the class $H$ of a line as generator. Since any two lines are linearly
equivalent, and since two distinct lines meet in one point, we have
$H.H=1$. This determines the intersection pairing on ${\mathbb P}^2$ by
linearity. The anticanonical class is 
$-K=3H$. Using the Nakai-Moishezon criterion, we can easily prove that
$-K$ is ample and we can define the corresponding  degree $d_{-K}$. 
The divisors D with a vanishing virtual genus and a degree
$d_{-K}(D)$, such as $0 \leq d_{-K}(D) \leq 9$, 
%(or $-1 \leq d_{-K}(D) \leq 10$) 
are given by $D_{M2}=H$ and $D_{M5}=2H$. 

Here we may generalize \cite{inv} and remark that the (anticanonical) 
degree is actually equal to  the degree of the potential differential form 
coupled to the corresponding brane \ie  its spacetime dimension.
In \cite{inv} this was noticed as a coincidence, here we deduce this fact 
from the generalized U-duality superalgebra of  \cite{cjlp2}.
Later on we shall take this for granted when trying to find the
algebras corresponding to other string models at least for some degree to be 
determined.
 The divisor $D_{M2}$ of degree 3 corresponds to the $M2$ brane
and $D_{M5}$ of degree 6 to the $M5$ brane.                
$D_{M2}$ and $D_{M5}$ verify the Serre duality equation
$D_{M2}+D_{M5}=-K$ which corresponds to the electric-magnetic
Hodge duality ($3+6=d-2=9$).  
We note that these divisors are effective and
have  vanishing genus. The correspondence between rational divisors of 
degree zero and BPS instantons (or rather the scalar fields to which they 
couple) was known for a long time \cite{man,manp}.

The duality of \cite{cjlp2} is itself demystified if we define a
Borcherds superalgebra by the simple fermionic superroot given by
$\alpha_0=H$. The Dynkin diagram, corresponding to the Cartan matrix   
$A=a=\left(\matrix {-1} \right) $ is given by:
\hspace{0.01mm} 
\begin{picture}(10,20)
\thicklines
\put(7,5){\circle{14}}
\put(8,-10){\makebox(0,0){$\alpha_0$}}
\put(7,5){\circle*{10}}
\end{picture}
\hspace{0.6cm}
= {\bf f}

\vspace{3mm}
We give here our conventions for Dynkin diagrams of Borcherds superalgebras, 
the last character is a chemical label that will eventually dispense us from 
writing diagrams. The number of bonds between simple roots is the opposite of 
the off diagonal element of the (symmetrized) Cartan matrix, it is in effect an 
intersection number and the roots come in (at least) four colors.
Imaginary roots are defined here as roots of norm $a_{ii} \leq
0$. Some of them do not generate infinite chains of roots, as in
the purely bosonic case. Indeed one can easily see from the Jacobi
identity that $ad(e_i)^2 e_j=0$ if $e_i$ is a fermionic root of null norm.

\begin{center}
\begin{tabular}{|l|l|l|}
\hline
\begin{picture}(14,14)
\thicklines
\put(7,4){\circle{14}}
\end{picture}
&
Bosonic real root of length 2
& {\bf B}
\\
\hline
\begin{picture}(14,14)
\thicklines
\put(7,4){\circle{14}}
\put(2,-1){\line(1,1){10}}\put(2,9){\line(1,-1){10}}
\end{picture}
&
Bosonic imaginary root of length $\leq$ 0
& {\bf b}
\\
\hline
\begin{picture}(14,14)
\thicklines
\put(7,4){\circle*{14}}
\end{picture}
&
Fermionic ``imaginary'' root of length 1, $a_{ii}=0$ 
&{\bf F}
\\
\hline
\begin{picture}(14,14)
\thicklines
\put(7,4){\circle{14}}
\put(7,4){\circle*{10}}
\end{picture}
&
Fermionic imaginary root of length $\leq$ -1 
& {\bf f}
\\
\hline
\end{tabular}
\end{center}

In the M-theory case, the positive superroots are fermionic and bosonic
$\pi_n=nH$ with $n=1$ or  $n=2$.
By truncating to the set of positive superroots  
we obtain the following superalgebra   
\be
\{e_{\alpha_0},e_{\alpha_0}\}= -e_{\pi_2}\ ,\qquad
{[}e_{\alpha_0},e_{\pi_2} {]}=0\ ,\qquad {[} e_{\pi_2}, e_{\pi_2}{]}
=0\ .
\label{d11com}
\ee
The truncation to the finite dimensional superalgebra of \cite{cjlp2}
is defined by a ${\mathbb Z}$-gradation and is thus 
consistent, one may see it as a two step process namely restriction to positive degrees followed by truncation to some maximal degree at our disposal. 

We have related three symmetries and the intersection form:
 the U-duality was known to be related to the part of the second
cohomology  orthogonal to the canonical class,  the rest of the cohomology  
contains the extended superduality of \cite{cjlp2} and corresponds actually to 
a Borcherds superalgebra of which the former is a truncation.
Using chemical nomenclature for Dynkin molecules we may call the M theory 
Borcherds algebra: {\bf f}. Actually in this case the superalgebra is finite 
dimensional and equal to $OSp(1|2)$

In \cite{cjlp2} the authors 
introduced a pseudo-involution $\CS$ that exchanges the
generators $e_{\pi_1}$ and $e_{\pi_2}$:
\be
\CS \, e_{\pi_N}  = \pm e_{-\pi_N-K} ,\qquad
\CS^2 \, e_{\pi_N} = \epsilon_N e_{\pi_N}
.\label{d11om}
\ee
(There is a global sign arbitrariness which has no meaning but the  relative
$\epsilon_N$ is 
there  to compensate for the square of Hodge dualisation  
that maybe equal to $-1$ for some values of the rank of the 
corresponding field strength $1+grad(\pi_N)$.)
The operator $\CS$ is well-defined even after truncation provided the spectrum 
of truncated positive superroots is invariant under Serre duality, this
requires the appropriate choice of maximal degree: the spacetime dimension 
minus two.
Note that here we have $\CS^2\, e_{\alpha_i} = -e_{\alpha_i}$, so 
$\CS^2 = - {\rm id}$. In lower dimensions we
shall see that $\CS^2$ acts sometimes as an involution, and sometimes
as a pseudo-involution. Let us note that $\CS$ does not
preserve the commutation relations (\ref{d11com}), up to signs it  
implements Serre duality.

Let us introduce the following nonlinear ``potential'' differential form: 
\be
\v = e^{A_\3\, e_{\alpha_0}}\, e^{\tA_\6\,e_{\pi_2} }\ .\label{d11coset}
\ee
The Grassmann {\it angle} $A_\3$ (resp. $\tA_\6$) is a 3-form (resp. 6-form) 
coupled to the M2- (resp. M6-) brane  and 
defined on an eleven dimensional manifold. This manifold should be describable 
on the algebraic side of the correspondence, presently we can read  from there 
its dimension $d$, the Hodge duality. We may conjecture that the spacetime 
coordinates will appear as algebraic moduli.  Note that the generators 
$e_{\alpha_i}$ are even or odd according to whether the degrees of the
associated field strengths are odd or even. The odd generator corresponds to 
what was called the 12th fermionic dimension in  \cite{cjlp2}. 

By an elementary calculation one checks that the field strength $\G=d\v\,
\v^{-1}$ following from (\ref{d11coset}) is given by
\bea
\G &=& dA_\3\,e_{\alpha_0}  + (d\tA_\6 - \ft12 A_\3 \wedge dA_\3)
\,e_{\pi_2} \ ,\nn\\
  &=& F_\4 \, e_{\alpha_0} + \tF_\7\, e_{\pi_2}\ .\label{gcalc}
\eea  
Note that when the exterior
derivative passes over a generator, the latter acquires a minus sign if it
is odd.  Thus $d(e_{\pi_i} \ A_n)=(-)^{grad(\pi_i)} e_{\pi_i}
\ dA_n$. We recall the (twisted) self-duality equation:
\be
{*\G} = \CS \, \G\ ,\label{d11master}
\ee
in components:  
\be
{*F_\4} = \tF_\7\equiv d\tA_\6 - \ft12 A_\3\wedge F_\4\ ,\label{d11fo}
\ee 
Since the
doubled field strength $\G$ is written as $\G= d\v\, \v^{-1}$, it
follows by taking an exterior derivative that we have the
Cartan-Maurer equation $d\G= -d\v \wedge d\v^{-1} = d\v\, \v^{-1} \wedge d\v\,
\v^{-1}$, and hence
\be
   d\G - \G\wedge \G = 0\ .\label{cm}
\ee
Now, substituting (\ref{d11fo}) into (\ref{cm}), 
it follows that
\be
d{*F_\4} + \ft12 F_\4\wedge F_\4 = 0\ .\label{d11eom}
\ee
This equation can be obtained by varying with respect to $A_{(3)}$ the
bosonic Lagrangian of eleven-dimensional supergravity \cite{cjs} given by
\be
{\cal L}_{11} = R\, {*\oneone} - \ft12 {*F_\4} \wedge F_\4 - \ft16
F_\4\wedge F_\4 \wedge A_\3\ ,\label{d11lag}
\ee

In \cite{llps} it has been shown that the superalgebra of gauge symmetries
(\ref{d11com}) implies nonlinear relations of the type 
$\, {t_{\pi_1} \over 2 \pi}.{t_{\pi_1} \over 2\pi}={t_{\pi_2}
\over 2\pi} \,$ where $t_{\pi_1}
$ ( resp. $t_{\pi_2} $) is 
the tension of the M2 brane (resp. M5 brane).
More generally, for each non-vanishing
commutator $[e_{\pi_i},e_{\pi_j}] \sim e_{\pi_i+\pi_j}$
 coming from a truncated Borcherds superalgebra, we can deduce a
 product relation between the tensions
$\, {t_{\pi_i} \over 2 \pi}.{t_{\pi_j} \over
2\pi}={t_{\pi_i+\pi_j} \over 2 \pi}$ . 
This correspondence has been recovered, on the del Pezzo side, in \cite{inv}. 
Applying the above rule gives also the
Dirac-Nepomechie-Teitelboim quantisation condition. For BPS states tensions
and  charges are proportional to each other.
These results have been reviewed in several Conference Proceedings 
to which we may refer \cite{proc}.
As the truncated
spectrum of positive roots is invariant by construction under the
Serre duality, for each $\pi_i$, $-K-\pi_i$ will be a positive
superroot. Another way to phrase this is to give a formula for $-K$ as a 
positive combination of simple roots.
%  when it is possible
Here $-K=3\alpha_0$.

\subsection{IIA theory}

Blowing up $\CP^2$ in one point, we get the del Pezzo surface
${\mathbb B}_1$. If $E_{11}$ is the exceptional divisor
corresponding to this blow up, the Picard lattice is now of rank two :
$Pic({\mathbb B}_{1})= {\mathbb Z}H+{\mathbb Z}E_{11}$. In this
orthogonal basis
the (unimodular) intersection matrix is
$\left(\matrix {1 & 0 \cr 
                 0 & -1}\right) \ .$

From ${\mathbb B}_1$, we may retrieve in the same way as above Type IIA
supergravity and its duality superalgebra as a truncated Borcherds algebra.
The anticanonical divisor is $-K=3H-E_{11}$. 
Divisors generating $Pic({\mathbb B}_1)$ are
$\alpha_0=H-E_{11}$ and $\alpha_1=E_{11}$.
Both have null  virtual genus and minimal positive degrees $-K.\alpha_i$.
In fact $-K=3\alpha_0+2\alpha_1$.

They give the intersection matrix
$A=\left(\matrix {0 & -1 \cr 
                 -1 & 1}\right)$
and the Cartan matrix 
$a=\left(\matrix {0 & -1 \cr
                 -1 & 0}\right)$
which satisfies Borcherds superalgebra axioms, with  a Dynkin diagram
given by:
\hspace{5mm}
\begin{picture}(42,20)
\thicklines
\put(0,5){\circle{14}}
\put(0,-10){\makebox(0,0){$\alpha_0$}}
\put(42,5){\circle*{14}}
\put(42,-10){\makebox(0,0){$\alpha_1$}}
\put(7,5){\line(1,0){28}}
\put(-5,0){\line(1,1){10}}\put(-5,10){\line(1,-1){10}}
\end{picture}
\hspace{5mm}
and its formula is {\bf bF}.

\vspace{3mm}
Chevalley-Serre-Kac-Moody-Borcherds relations allow us to construct
the positive roots of this algebra,
and by truncating to positive degrees lower than 9, we get:

\begin{center}
\begin{tabular}{|c|c|c|c|}
\hline
Degree & positive root & BPS state & potential \\
\hline
0 & 0& ?&$\phi$\\
1 & $\alpha_1$ & D0 & ${\cal A}_{(1)}$\\
2 & $\alpha_0$ & F1 & ${\cal A}_{(2)}$\\
3 & $\alpha_0 + \alpha_1$ & D2 & ${\cal A}_{(3)}$\\
5 & $2 \alpha_0 + \alpha_1$ & D4 & $\tA_{(5)}$\\
6 & $2 \alpha_0 + 2 \alpha_1$ & NS5 & $\tA_{(6)}$\\
7 & $3 \alpha_0 + \alpha_1$ & D6 & $\tA_{(7)}$\\
8 & $-K = 3 \alpha_0 + 2 \alpha_1$ &? & $\psi$ \\
\hline
9 & $4 \alpha_0 + \alpha_1$ & D8 & none \\
\hline
\end{tabular}
\end{center}

We remark that as far as the superalgebra structure is concerned
 we should actually choose maximal degree 8
and  preserve symmetry under Serre duality $C \ra -K-C$, 
%(except D8) 
this allows us to define the pseudo-involution
(\ref{d11om})  on the truncated positive superroot set. 
%(except $4 \alpha_0 + \alpha_1$). 
It corresponds on the associated forms to Hodge duality
\be
d {\cal A}_{(i)} = \pm \ast d \tA_{(8-i)} \ + {\rm nonlinear \, terms.} 
\ee
As there is no dynamical field coupled to the D8 brane it stays alone, but note that it is rational and secondly that it has no Serre dual in the root system.
In other words rationality as we have seen is a Serre duality invariant concept
whereas the existence of BPS states as we define them today is not.

We also note that most of these positive roots are rational curves (including 
the root corresponding to the D8 brane) and
correspond to $\frac{1}{2}$-BPS states whose brane spacetime dimension
is given 
by the degree, the only non rational curve is simply the Serre dual of zero,
in other words the anticanonical divisor.
Adding to the algebra generators $e_{\alpha_0}$ and $e_{\alpha_1}$ the
Cartan element $h={1 \over 2}h_{\alpha_0}-h_{\alpha_1}$ and dropping the other 
Cartan generator as in M theory we get the superalgebra relations:

\be
 \begin{array}{l l l}
{[} h, e_{\alpha_1} {]} = -\ft32 e_{\alpha_1}\ , & 
{[} h, e_{\alpha_0} {]} =        e_{\alpha_0}\ , &
{[} h, e_{\alpha_1+\alpha_0} {]} = -\ft12 e_{\alpha_1+\alpha_0} \ ,
\\
{[} h, e_{3\alpha_0+\alpha_1} {]} = \ft32  e_{3\alpha_0+\alpha_1}  \ , &
{[} h,  e_{2\alpha_0+2\alpha_1} {]} = - e_{2\alpha_0+2\alpha_1} \ , &
{[} h,  e_{2\alpha_0+\alpha_1} {]} = \ft12 e_{2\alpha_0+\alpha_1}  \
. 
\end{array}
\label{2ahcom}
\ee
\be
\begin{array}{l l l}
{[} e_{\alpha_1} , e_{\alpha_0} {]} = -e_{\alpha_1+\alpha_0} \ , &
\{ e_{\alpha_1} , e_{\alpha_1+2\alpha_0} \} = -
e_{2\alpha_1+2\alpha_0} \ , &
{[} e_{\alpha_0} , e_{\alpha_1+\alpha_0} {]} = -
e_{\alpha_1+2\alpha_0} \ ,
\\
{[} e_{\alpha_0} , e_{\alpha_1+2\alpha_0} {]} = -
e_{\alpha_1+3\alpha_0} \ , &
 \{ e_{\alpha_1+\alpha_0} ,
e_{\alpha_1+\alpha_0} \} = - e_{2\alpha_1+2\alpha_0} \ . & 
\end{array}
\label{2acom}
\ee
\be
\begin{array}{l l}
\{ e_{\alpha_1}, e_{\alpha_1+3\alpha_0} \}= \ft38
e_{2\alpha_1+3\alpha_0} \ , & 
{[} e_{\alpha_0}, e_{2\alpha_1+2\alpha_0} {]} =\ft28
e_{2\alpha_1+3\alpha_0} \ , \\
\{ e_{\alpha_1+\alpha_0},e_{\alpha_1+2\alpha_0} \} = \ft18
e_{2\alpha_1+3\alpha_0} \ . & 
\end{array}
\label{2athcom}
\ee

As above, one can associate to this superalgebra the potential
\be
\v = e^{\fft12\phi h}\, e^{{\cal A}_\1 e_{\alpha_1}}\, e^{{\cal A}_\2
e_{\alpha_0}}\, e^{{\cal A}_\3 e_{\alpha_1+\alpha_0}}\,
e^{\tA_\5 e_{\alpha_1+2\alpha_0}}\, e^{\tA_\6 e_{2\alpha_1+2\alpha_0}}\, e^{\wtd{\cal A}_\7 e_{\alpha_1+3\alpha_0}}\,
e^{\fft12\psi e_{2\alpha_1+3\alpha_0}}\ .
\ee
and derive its field strength $\G= d\v\, \v^{-1}$. Let us comment 
that as in the previous case and in the sequel we do not see in the field
theory description (even after doubling) any field for one of  the 
Cartan generators, in the present case we use only the combination 
$h={1 \over 2}h_{\alpha_0}-h_{\alpha_1}$ for instance). We are dropping the 
($-K$)-degrees above 8 and hope to return to this fact in a later work. 
Nevertheless it corresponds to a trivial scaling symmetry of the field 
equations.

The Cartan-Maurer equation (\ref{cm}), combined with the self-duality
condition (\ref{d11master}), gives precisely the equations of motion
for the bosonic part of the IIA 10-dimensional supergravity lagrangian
\bea
{\cal L}_{10} &=& R {*\oneone} -\ft12 {*d\phi}\wedge d\phi - \ft12
H
e^{-\fft32\phi}\, {*{\cal F}_\2}\wedge {\cal F}_\2 -\ft12 e^{\phi}\,
{*F_\3}\wedge F_\3\nn\\
&& - \ft12 e^{-\fft12\phi}\, {*F_\4}\wedge F_\4 -\ft12
dA_\3\wedge dA_\3\wedge A_\2\ ,
\eea
where $F_\4=dA_\3 -dA_\2\wedge {\cal A}_\1$, $F_\3=dA_\2$ and ${\cal
F}_\2= d{\cal A}_\1$.

We have checked the roots corresponding to the fields of IIA SUGRA by
using the denominator formula (see for instance \cite{ray}).
We shall complete the same checks for IIB theory and M theory on $T^2$ below.

\subsection{IIB theory}

We are now ready for the most striking evidence for a Borcherds algebra
in string theory. The generators are all bosonic (only even differential forms 
appear).
The unique other smooth del Pezzo surface with D=10 is $\CP^1 \times
\CP^1$ whose Picard group is generated by the two classes of $\CP^1$'s
$l_1$ and $l_2$. If we see $\CP^1 \times
\CP^1$ as a trivial bundle with fiber $l_1$ and section $l_2$, we
deduce that $l_1^2 = 0$ as two fibers do not intersect (similarly
$l_2^2=0)$ and $l_1.l_2=1$ as the section and one fiber intersect in
one point. Following the line of what we have just done for ${\mathbb B}_1$,
this surface will now correspond to type IIB 10-dimensional
supergravity. 
Divisors generating $Pic({\mathbb F }_0)$ are
$\alpha_0=l_1$ and $\alpha_1=l_2-l_1$.
Here $-K=4\alpha_0 +2\alpha_1$.

The Cartan matrix $a_{ij}=A_{ij}$ is $\left(\matrix {0 & -1 \cr 
                 -1 & 2}\right)$.

Its associated Borcherds algebra  has one lightlike simple root, it
plays a crucial role in the  construction of the Monster algebra
\cite{bor2} and has been analyzed in \cite{sla}.
The (purely bosonic) Dynkin diagram is represented by:
\hspace{5mm}
\begin{picture}(42,20)
\thicklines
\multiput(0,5)(42,0){2}{\circle{14}}
\put(0,-10){\makebox(0,0){$\alpha_0$}}
\put(42,-10){\makebox(0,0){$\alpha_1$}}
\put(7,5){\line(1,0){28}}
\put(-5,0){\line(1,1){10}}\put(-5,10){\line(1,-1){10}}
\end{picture}
\vspace{0.3cm}\hspace{5mm}
with formula {\bf bB}.
Let us recall that the node $\alpha_1$ corresponds to the S-duality symmetry
of IIB theory, which correponds to the permutation of the $\CP^1$'s, 
it is different from the IIA node with the same label, in fact 
they will span two of the three dimensions of the reduction to 9 dimensions. 
$\alpha_1$ spans  the part of the cohomology that is orthogonal to
the canonical class $K$. Let us note also that we have broken the manifest 
(Weyl) symmetry between the two $\CP^1$'s but only in the choice of simple 
roots.
 
The positive roots, of degree lower than 10, are
\begin{center}
\begin{tabular}{|c|c|c|c|}
\hline
Degree & positive root & BPS state & $d$-form \\
\hline
0 & 0 & ? & $\phi$\\
0 & $\alpha_1$ & D-1 & $\chi$\\
2 & $\alpha_0$ & F1 & $A_\2^2$\\
2 & $\alpha_0 + \alpha_1$ & D1 & $A_\2^1$\\
4 & $2 \alpha_0 + \alpha_1$ & D3 & $B_\4$\\
6 & $3 \alpha_0 + \alpha_1$ & D5 & $\tA_\6^1$\\
6 & $3 \alpha_0 + 2\alpha_1$ & NS5 & $\tA_\6^2$\\
8 & $ 4 \alpha_0 +  \alpha_1$ &D7 & $\wtd\chi$ \\
8 & $-K= 4 \alpha_0 + 2\alpha_1$ & ? & $\psi$\\
\hline
8 & $4 \alpha_0 + 3\alpha_1$ & NS7 & none \\
\hline
\end{tabular}
\end{center}

It should be noted that the divisor $-\alpha_1$ which is a negative root
has vanishing virtual genus and  degree zero. It corresponds
to the NS-1 instanton which is S-dual to D-1 and completes the triplet of 
sources for the 3 fields of the ($SO(2)$) gauge invariant formulation
of the scalar sector. 
We can now introduce the formal form $\v$ given by    
\bea
\v &=& e^{\fft12\phi\, h}\, e^{\chi\, e_{\alpha_1}}\, e^{(A_\2^1 \, 
e_{(\alpha_0+\alpha_1)} + 
A_\2^2\, e_{\alpha_0})}\, e^{B_\4\, e_{(2\alpha_0+\alpha_1)}}\,
\nn\\
&&e^{(\tA_\6^1\, e_{(3 \alpha_0 + \alpha_1)} + \tA_\6^2\, 
e_{(3 \alpha_0 + 2\alpha_1)}
)}\, e^{\wtd\chi\, e_{(4\alpha_0 + 2\alpha_1)}}\, e^{\fft12\psi\, 
e_{(4\alpha_0+3\alpha_1)}}\ .
%\label{2bcoset}
\eea
where $h=h_{\alpha_1}$.

One can define again an involution (\ref{d11om}) such that the $2^{\mathrm nd}$ 
order equations derived from (\ref{cm}) and (\ref{d11master}) come from the
bosonic part of the IIB 10-dimensional supergravity Lagrangian. The  
positive root  $-K$ has multiplicity one not two, it is Serre dual to 
the zero ``root'' with generator $h$.

\subsection{M theory on $T^k , k \leq  8$}

We consider ${\mathbb B}_k$ obtained by blowing up $k$ points in generic
positions on ${\mathbb P}^2$.
First, we will define a Borcherds
superalgebra associated to the Picard lattice. 
The truncation to positive
superroots whose degree is lower than $(9-k)$ will generate a superalgebra
invariant under Serre duality as a linear space. 
This will permit us to define a
pseudo-involution (\ref{d11om}). The Maurer-Cartan equation (\ref{cm})
combined with the self-duality condition (\ref{d11master}) will
reproduce the equations of motion of M-theory compactified on a $k$-dimensional
torus $T^k$. We note that the 3-dimensional 
Picard lattices of ${\mathbb B}_2$ and
$\CP^1 \times \CP^1$ blown up on a point are Lorentzian, self-dual and
odd. Consequently they are isomorphic, define the same Borcherds
superalgebra and map to the same supergravity theory. This corresponds
to the T-duality between $IIA$ and $IIB$ compactified on a circle.
     
The Picard lattice of ${\mathbb B}_k$ is generated by H (the class of a line in
${\mathbb P}^2$) and the exceptional divisors 
$E_{(11-i)} ,\ 0 \leq i \leq (k-1)$. We choose an appropriate basis
$(\alpha_i,\ 0 \leq i \leq (k-2), \beta,\gamma)$ for this lattice which 
corresponds to  the simple
superroots of a Borcherds superalgebra. Below, we list the simple superroots
 with the associated Dynkin diagrams for all $k$'s.

\begin{center}
\begin{tabular}{|c|c|c|c|}
\hline
$k$ & Simple superroot & Dynkin diagram & Formula\\
\hline
0 & $\begin{array}{cc}
 \beta=& H       
 \end{array}$ &
\begin{picture}(10,20)
\thicklines
\put(7,5){\circle{14}}
\put(8,-10){\makebox(0,0){$\beta$}}
\put(7,5){\circle*{10}}
\end{picture}
& {\bf f}
\\
1 & $\begin{array}{cc}
  \beta=&H-E_{11}   \\
 \gamma=&E_{11}   
\end{array}$ &\begin{picture}(42,40)
\thicklines
\put(0,5){\circle{14}}
\put(0,-10){\makebox(0,0){$\beta$}}
\put(42,5){\circle*{14}}
\put(42,-10){\makebox(0,0){$\gamma$}}
\put(7,5){\line(1,0){28}}
\put(-5,0){\line(1,1){10}}\put(-5,10){\line(1,-1){10}}
\end{picture} 
&{\bf bF} \\
2 &   $\begin{array}{cc}
 \alpha_0=& E_{11}-E_{10}       \\
 \beta=&H-E_{11}-E_{10}   \\
 \gamma=&E_{10}   
\end{array}$    & 
\begin{picture}(42,40)
\thicklines
\put(0,-20){\circle{14}}
\put(0,-35){\makebox(0,0){$\alpha_0$}}
\put(42,-20){\circle*{14}}
\put(42,-35){\makebox(0,0){$\gamma$}}
\put(42,22){\circle*{14}}
\put(54,22){\makebox(0,0){$\beta$}}
\put(7,-20){\line(1,0){28}}
\put(42,-13){\line(0,1){28}}
\end{picture}
&{\bf BFF}
\\
3 &$\begin{array}{cc}
\alpha_{0}=&E_{11}-E_{10} \\
\alpha_{1}=&E_{10}-E_{9}   \\
\beta=&H-E_{11}-E_{10}-E_{9} \\
\gamma=&E_{9} 
\end{array}$ &
\begin{picture}(84,60)
\thicklines
\multiput(0,-20)(42,0){2}{\circle{14}}
\put(0,-35){\makebox(0,0){$\alpha_0$}}
\put(42,-35){\makebox(0,0){$\alpha_1$}}
\put(84,-20){\circle*{14}}
\put(84,-35){\makebox(0,0){$\gamma$}}
\put(84,22){\circle{14}}
\put(100,22){\makebox(0,0){$\beta$}}
\put(7,-20){\line(1,0){28}}
\put(49,-20){\line(1,0){28}}
\put(84,-13){\line(0,1){28}}
\end{picture}
&{\bf BBFB}
 \\
$4$ to $8$ &

$\begin{array}{cc}
\alpha_{i}=&E_{11-i}-E_{10-i} ,\\
& 0 \leq i \leq (k-2) \\
\beta=&H-E_{11}-E_{10}-E_{9}   \\
\gamma=&E_{12-k}\\
\\ 
\end{array}$
& 
\hspace{1mm}
\begin{picture}(140,60)
\thicklines
\multiput(0,-20)(42,0){4}{\circle{14}}
\put(0,-35){\makebox(0,0){$\alpha_0$}}
\put(42,-35){\makebox(0,0){$\alpha_1$}}
\put(84,-35){\makebox(0,0){$\alpha_2$}}
\put(126,-35){\makebox(0,0){$\alpha_{(k-2)}$}}
\put(126,22){\circle*{14}}
\put(142,22){\makebox(0,0){$\gamma$}}
\put(84,22){\circle{14}}
\put(100,22){\makebox(0,0){$\beta$}}
\put(7,-20){\line(1,0){28}}
\put(49,-20){\line(1,0){28}}
%\put(133,-20){\line(1,0){28}}
\put(91,-20){\dashbox{3}(28,0)}
\put(84,-13){\line(0,1){28}}
\put(126,-13){\line(0,1){28}}
\end{picture}

\hspace{1mm}

&{\bf BB(BB)B.BF}
\\
&&&
\\
\hline
\end{tabular}
\end{center}

The roots $(\alpha_i,\gamma)$ define a sl($k \arrowvert 1$)
superalgebra and the
roots $(\alpha_i,\beta)$ represented by instantons, \ie divisors with 
vanishing degree and virtual genus, define the Dynkin
diagram of the U-duality group $E_{k}$. 

The positive roots, of degree lower than $9-k$ 
($3\le k \le 8$) whenever they exist are 
\begin{center}
\begin{tabular}{|c|c|c|c|}
\hline
Degree & positive root & BPS state & $d$-form \\
\hline
0 & 0 & dilatons &$\vec\phi$  \\
0 & $H-E_i-E_j-E_l$ & thrice-wrapped M2 & $A_{\0 ijl}$\\
0 & $E_i-E_j i > j$ & Kaluza-Klein modes & ${\cal A}^i_{\0 j}$\\
1 & $E_i$ & momentum & $\hA_\1^i$\\
1 & $H-E_i-E_j$ & twice-wrapped M2 & $ A_{\1 ij}$\\
2 & $H-E_i$ & once-wrapped M2 & $A_{\2 i}$\\
3 & $H$ & M2 & $A_\3$\\
$6-k$ & $2H-\sum_p E_p$ & $(k)$-wrapped M5 & $\tA_{(6-k)}$\\
$7-k$ & $2H-\sum_{p \neq i} E_p$ & $(k-1)$-wrapped M5& $\tA_{(7-k)}^i$ \\
$8-k$ & $2H-\sum_{p \neq i \, j} E_p$ & $(k-2)$-wrapped M5
&$\tA_{(8-k)}^{ij}$  \\
$8-k$ & $3H-\sum_{p \neq i} E_p-2E_i$ & magnetic dual of momentum & $\tcA_{(8-k)i}$  \\
$9-k$ & $3H-\sum_{p \neq i \, j } E_p -2E_i$ & magnetic dual of Kaluza-Klein & $\tcA_{(9-k)i}^j$  \\
$9-k$ & $2H-\sum_{p \neq i \, j \, l} E_p$ & $(k-3)$-wrapped M5 & $\tcA_{(9-k)}^{ijl}$  \\
$9-k$ & $-K=3H-\sum_p E_p$ & magnetic dual of dilatons & $\vec\psi$ \\
\hline
\end{tabular}
\end{center}
We have included the dilatons $\vec\phi$. 
These truncations define precisely the superalgebras which appeared in 
\cite{cjlp2}.
Defining the involution (\ref{d11om}), the $2^{\mathrm nd}$ order equations
derived from (\ref{cm}) and (\ref{d11master}) come from the
bosonic part of the  11-dimensional supergravity lagrangian
compactified on the $k$-dimensional torus $T^k$. We shall return to the 
denominator formula in a subsequent paper, for the time being 
we have mostly done random checks of the root structure beyond $k=3$. 
  
By construction all the intersection matrices are unimodular. 
Let us notice  that the case of spacetime dimension 3
and Picard rank 9 has the odd simple root at the location of the
affine root of $E_9$.
Indeed $-K$ can be expressed as the sum  of the most positive root of
$E_8$ with this odd simple root, it is orthogonal to all roots of $E_8$. 
In fact this is probably  true in all cases for $D=3$. (We have checked it for
all cases where the U-duality algebra is a simple Lie algebra.)
 
Now, one may consider ${\mathbb B}_9$, it is not a del Pezzo surface as 
$K^2_{{\mathbb B}_9}=0$ instead $K_{{\mathbb B}_9}$ is ``nef'' (\ie 
$K_{{\mathbb B}_9}.e \ge 0$ 
for any effective divisor on the surface). 
In an
appropriate basis \cite{vaf}, the Picard lattice  
$Pic({\mathbb B}_9)$ has intersection form
$$-\Gamma_{E_8}+\left(\matrix {1 & 0 \cr 
0 & -1}\right)$$  with $\Gamma_{E_8}$ the Cartan matrix of the
exceptional Lie algebra  $E_8$. 
The formula for its diagram is ${\bf B_2(BB)B_5F}$, it comes from a
unimodular lattice of (maximal) rank 10 to be compared to the
even unimodular lattice $E_{10}$: ${\bf B_2(BB)B_6}$.
\vfill
\eject
\section{Type I / Heterotic}

We know that 10-dimensional Type IIA and IIB theories correspond to smooth del
Pezzo surfaces ${\mathbb B}^1$ and $\CP^1 \times \CP^1$.  Is there
any surface whose associated Borcherds algebra gives type I or
Heterotic theories ? Our answer is that it is possible to retrieve the 
truncated
version of these theories where the gauge sector has been dropped in such
a way that their massless sector is type I supergravity.

\subsection{Physical requirements}

It is well known that Type I and Heterotic theories, deprived of their
gauge sector and compactified on tori $T^k$, $k\ge 3$,
have U-duality Lie groups $A_1 \times A_1$, $A_3$, $D_4$, $D_5$, $D_6
\times A_1$, $D_8$ instead of the $E_k$
sequence of Type II.

The 10-dimensional Type I theory has as BPS states the D1 and D5 branes, while
Heterotic theories have the fundamental string F1 and the NS5 brane.
When
compactifying on tori, BPS states corresponding to those objects
may be
wrapped on tori. Instantons correspond to the simple roots of
the U-duality algebra, and one sees that the $A_1$ root which appears
in 8 dimensions comes from a twice wrapped 2-brane whereas the $A_1$
root appearing in 4 dimensions comes from a fully wrapped 5-brane. 
This is shown in the following Lie Dynkin diagram, where simple roots 
involving $E_i$ progressively appear in dimensions lower than $i$.

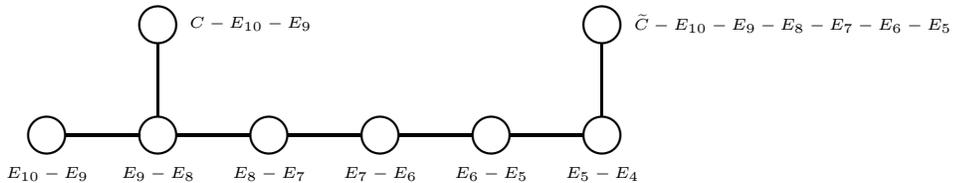
\begin{center}
\vspace{5mm}
\hspace{10mm}
\begin{picture}(280,60)
\thicklines
\multiput(0,0)(42,0){6}{\circle{14}}
\put(210,42){\circle{14}}
\put(42,42){\circle{14}}
\put(0,-15){\makebox(0.4,0.6){\tiny $E_{10}-E_9$}}
\put(42,-15){\makebox(0.4,0.6){\tiny $E_9-E_8$}}
\put(84,-15){\makebox(0.4,0.6){\tiny $E_8-E_7$}}
\put(126,-15){\makebox(0.4,0.6){\tiny $E_7-E_6$}}
\put(168,-15){\makebox(0.4,0.6){\tiny $E_6-E_5$}}
\put(210,-15){\makebox(0.4,0.6){\tiny $E_5-E_4$}}
\put(77,42){\makebox(0.4,0.6){\tiny $C-E_{10}-E_9$}}
\put(282,42){\makebox(0.4,0.6){\tiny $\tC-E_{10}-E_9-E_8-E_7-E_6-E_5$}}
\put(7,0){\line(1,0){28}}
\put(49,0){\line(1,0){28}}
\put(91,0){\line(1,0){28}}
\put(133,0){\line(1,0){28}}
\put(175,0){\line(1,0){28}}
\put(42,7){\line(0,1){28}}
\put(210,7){\line(0,1){28}}
\end{picture}
\hspace{10mm}
\vspace{5mm}
\end{center}

We deal here with the type I case, but the story is the same for
the Heterotic case, if one exchanges D1 and D5 for F1 and NS5. 
Writing $C$ and $\tC$ for the rational divisors corresponding to D1
and D5 branes, the BPS states and their corresponding rational
divisors for the $k^{\mathrm th}$ toroidal compactification are presented in the
following table, where the $E_i$'s are the exceptional $(-1)$-curves
associated with the successive compactifications.

\begin{center}
\begin{tabular}{|c|c|c|c|}
\hline
Degree & positive root & BPS state & $d$-form \\
\hline
0 & 0 & dilatons &$\vec\phi$ \\
0 & $C-E_i-E_j$ & twice-wrapped D1 & $A_{\0 ij}$\\
0 & $E_i-E_j$ & Kaluza-Klein modes & ${\cal A}^i_{\0 j}$\\
1 & $E_i$ & momentum & $\hA_\1^i$\\
1 & $C-E_i$ & once-wrapped D1 & $ A_{\1 i}$\\
2 & $C$ & D1 & $A_{\2}$\\
$6-k$ & $\tC-\sum_p E_p$ & $(k)$-wrapped D5 & $\tA_{(6-k)}$\\
$7-k$ & $\tC-\sum_{p \neq i} E_p$ & $(k-1)$-wrapped D5& $\tA_{(7-k)}^i$ \\
$7-k$ & $-K-E_i$ & magnetic dual of momentum & $\tcA_{(7-k)i}$ \\
$8-k$ & $-K-E_i+E_j$ & magnetic dual of Kaluza-Klein & $\tcA_{(8-k)i}^j$\\
$8-k$ & $\tC-\sum_{p \neq i \, j} E_p$ & $(k-2)$-wrapped D5
&$\tA_{(8-k)}^{ij}$  
\\
$8-k$ & $-K$ & magnetic dual of dilatons& $\vec\psi$ \\
\hline
\end{tabular}
\end{center}

\subsection{Ampleness of $-K$}
 
We would like now to generalize the correspondence between algebraic surfaces
and string theories from M-theory to the 16 supercharges case. Let us try at 
first to work with any ample divisor. It will be needed to define a degree and 
a projective embedding that guarantees the isomorphism between Cartier classes 
and the Picard group.   
The degree of a divisor is given by its scalar 
product with the ample divisor $H_k$.
For a divisor corresponding to a BPS state we shall assume it is its physical
spacetime
dimension. In particular, we have $H_0.C=2$. From  the degrees of $E_i$
and $E_i-E_j$ we  deduce that
when we blow up the $k^{\mathrm th}$ point, we have
$H_k=H_{k-1}-E_{11-k}$. We also know from general properties of the blowing-up
operation that $K_k=K_{k-1}+E_{11-k}$.

$-K_k$ must be Cartier of degree $8-k=D-2$ because we assume Serre duality 
which corresponds in the physical space-time to the Hodge duality between
$p$-forms and $(D-2-p)$-forms, this is by definition the 
Gorenstein property. The D1 and D5 branes being
Hodge duals in 10 dimension, we must take $\tC=-K_0-C$.

As the $E_i$'s are exceptional curves that appear when blowing up, they
are orthogonal  to each other and to $C$ and $\tC$. $C-E_{10}-E_9$
is a Lie simple root, it must be of self-intersection $-2$ like
$E_i-E_j$. It follows that $C^2=0$. From the Lie root 
$\tC-E_{10}-E_9-E_8-E_7-E_6-E_5$ we get similarly $\tC^2=4$.
 From the Adjunction formula
applied to the rational divisors $C$ and $\tC$ we now get $C.K_0=-2$
and $\tC.K_0=-6$ by assuming these divisors are still rational.

In summary, we must have
\bea
H_0.C=2=-K_0.C\\
H_0.\tC=6=-K_0.\tC
\eea

The moduli space of 10-dimensional Type I theory has dimension 2, and so must 
the Picard lattice of our surface X. We see from $C^2=0$ and 
$\tC^2=4$ that $C$ and $\tC$ are linearly independent, and we 
can conclude that $H_0=-K_0$ in Pic(X), which gives $H_k=-K_k$ for
compactifications. In particular the anticanonical classes $-K_k$ must 
be ample, which leads to the important conclusion that these surfaces 
must be del Pezzo.

\subsection{Finding the surface}

Du Val \cite{duv}, Demazure \cite{dem} and Hidaka-Watanabe \cite{hid} 
classified all Picard rank 2, normal, Gorenstein del Pezzo's,
which are in finite number. In particular for $D=K^2+2=10$ there are exactly 
three such surfaces, none of them having a Picard group corresponding to 
(truncated) Type I (or Heterotic) theory. It is important to note that
the algebraic surface we are looking for should not have any other rational 
curve of degree between 0
and 10 than those listed in the above table, which excludes $\CP^1
\times \CP^1$ which satisfies all the other requirements.

So if our surface exists, it must be a 
nonnormal (Gorenstein) del Pezzo. We recall that {\it nonnormal} means
that there is a line singularity, and {\it Gorenstein} means that the 
anticanonical class $-K$ exists as a Cartier divisor.

One finds in Miles Reid's classification of Gorenstein nonnormal del Pezzo's
\cite{rei1} two surfaces satisfying all the desired properties.  Let
us describe the first one. (The second is a kind of
degenerate case of it and has the same Picard group.) 

The rational scroll $Y := {\mathbb F}_{a;b}$  \cite{rei2} 
is defined as ${\mathbb C}^2
\times {\mathbb C}^2$ modded out by the two equivalence relations
\bea
(t_1,t_2,x_1,x_2) &\sim& (t_1,t_2,\mu x_1,\mu x_2) \\
(t_1,t_2,x_1,x_2) &\sim& (\lambda t_1, \lambda t_2, \lambda^{-a-b} x_1,
\lambda^{-b} x_2) \ .
\eea
It is clear that the first relation makes ${\mathbb F}_{a;b}$  a
$\CP^1$-bundle. Its base is also isomorphic to a $\CP^1$, the
projection being defined by the
ratio $t_1:t_2$. It can be embedded in $\CP^{a+2b+1}$ in the following way:
\be
(t_1,t_2,x_1,x_2) \mapsto (t_1^{a+b} \, x_1,t_1^{a+b-1} t_2 \, x_1,
\cdots, t_2^{a+b} \, x_1, t_1^b \, x_2, t_1^{b-1} t_2 \, x_2,
\cdots,t_2^b \, x_2)
\ .
\ee

The Picard group of the surface $Y$ is generated by two irreducible divisor 
classes: the fiber $A$, of self-intersection 0 and a $(-a)$-curve $B$,
defined by $x_1=0$, which is a section of the bundle 
and therefore has intersection 1 with $A$.

Let us now take $a=4$ and $b=2$. $B$ defines a plane in $\CP^9$ described by
the vanishing of all coordinates but the last three. Then
let us pick a point in 
this plane which is not on $B$: $(0,0,0,0,0,0,0,0,1,0)$. Projecting $\CP^9$ 
on $\CP^8$ from this point, the image $X$ of ${\mathbb F}_{4;2}$ in $\CP^8$ 
is given by
\be
(t_1,t_2,x_1,x_2) \mapsto (t_1^6 \, x_1,t_1^5 t_2 \, x_1,
\cdots, t_2^6 \, x_1, t_1^2 \, x_2, t_2^2 \, x_2)
\ .
\ee

Our surface $X$ is isomorphic to ${\mathbb F}_{4;2}$ except for the conic 
$B$ which is mapped to a double line. Still denoting by $A$ and $B$ the images
of these curves in $X$, the Cartier divisors of X are generated by $B$ and $2A$. A
divisor is Cartier if it can locally be described as the intersection
of the surface with an hyperplane in $\CP^n$, and one sees that the
intersection of the surface with an hyperplane transverse to the
double locus contains two
fibers $A$, so that $A$ is not Cartier but $2A$ is.

The intersection of Cartier divisors of X 
is defined via its normalisation the normal 
surface ${\mathbb F}_{4;2}$. The anticanonical class is $-K_X=6A+B$ and 
is ample. Rational divisors (defined as divisors of vanishing 
virtual genus) of nonnegative degree lower than 10 are $C=2A$ and
$\tC=B+4A$ which 
correspond respectively to the D1 and D5 branes, and one can easily verify
that they satisfy all required properties. In particular, blowing up points 
in generic positions gives the surfaces corresponding to toroidal 
compactifications of this truncated Type I theory.

\subsection{Borcherds algebras}

We get from the surface just described the Borcherds algebras
of these compactifications, which are
described by the following Dynkin diagrams. Root lengths are easy
to calculate, if one remembers that $C^2=0$, 
$\tC^2=4$,
$C.\tC=2$,
$E_i^2=-1$, 
$E_i.E_j=0$  if  $i \neq j$ and
$E_i.C=E_j.\tC=0$.
In dimension 10 the Cartan matrix restricted to CaCl(X) is the non unimodular
but even and bosonic:
$$\left(\matrix {0 & -2 \cr
-2  & -4 }\right) \ .$$
 We find the following Dynkin diagrams:\\ 
D=10 
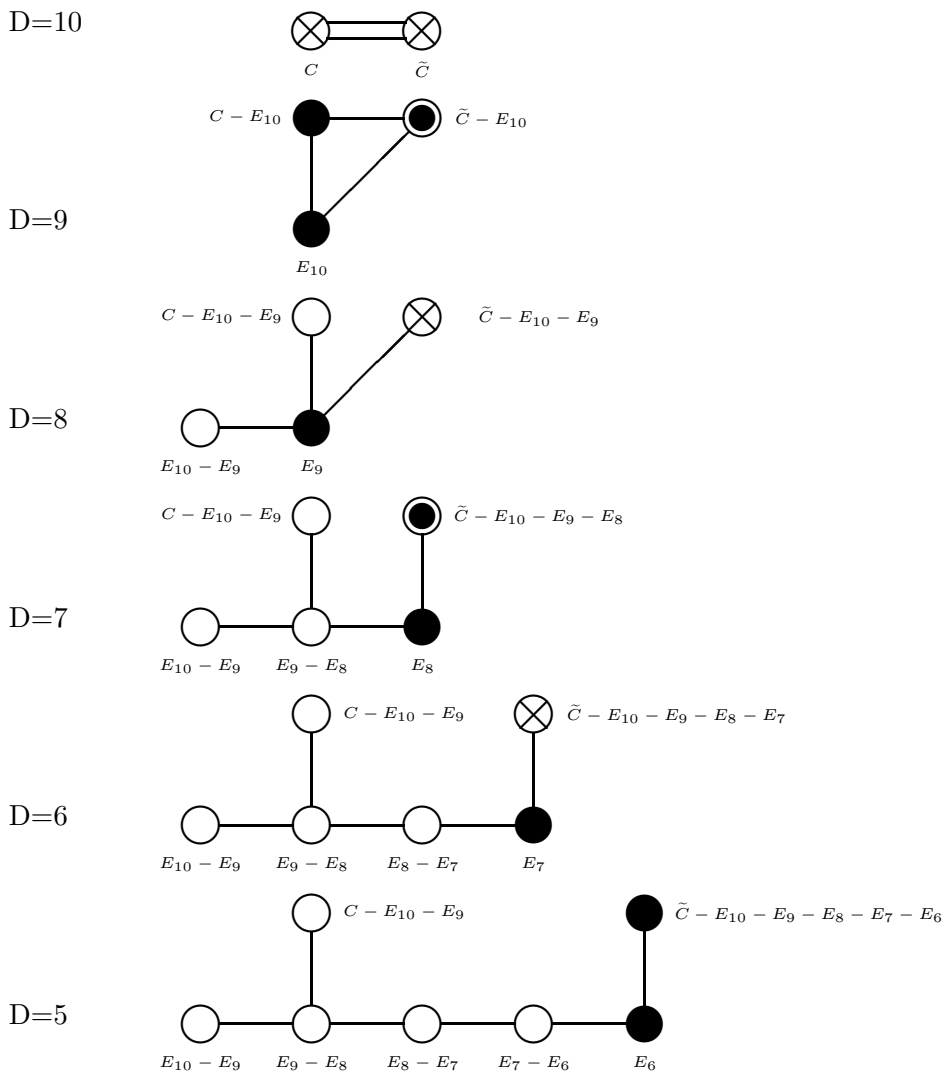
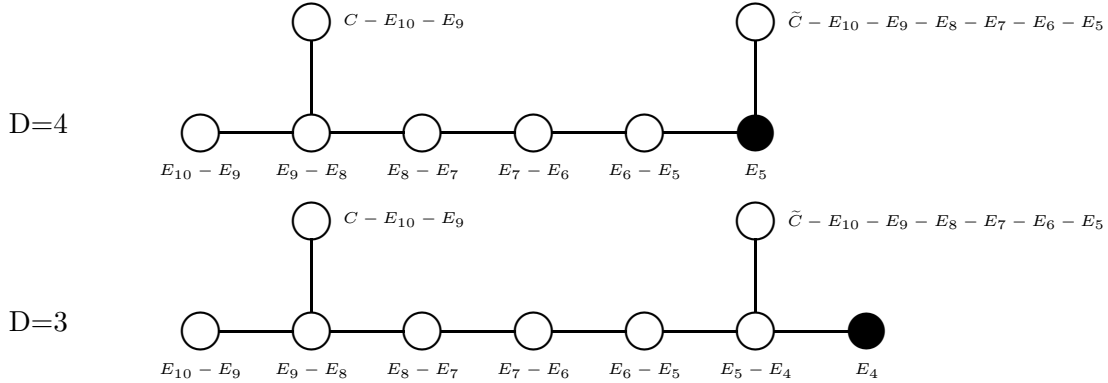
\hspace{13mm}
\begin{picture}(280,20)
\thicklines
\multiput(42,0)(42,0){2}{\circle{14}}
\put(42,-15){\makebox(0.4,0.6){\tiny $C$}}
\put(84,-15){\makebox(0.4,0.6){\tiny $\tC$}}
\put(48,-3){\line(1,0){30}}
\put(48,3){\line(1,0){30}}
\put(37,-5){\line(1,1){10}}\put(37,5){\line(1,-1){10}}
\put(79,-5){\line(1,1){10}}\put(79,5){\line(1,-1){10}}
\end{picture}
\vspace{5mm}
\\
D=9 
\vspace{5mm}
\hspace{15mm}
\begin{picture}(280,60)
\thicklines
\put(42,0){\circle*{14}}
\put(84,42){\circle{14}}
\put(84,42){\circle*{10}}
\put(42,42){\circle*{14}}
\put(42,-15){\makebox(0.4,0.6){\tiny $E_{10}$}}
\put(17,42){\makebox(0.4,0.6){\tiny $C-E_{10}$}}
\put(110,42){\makebox(0.4,0.6){\tiny $\tC-E_{10}$}}
\put(42,7){\line(0,1){28}}
\put(47,5){\line(1,1){32}}
\put(49,42){\line(1,0){28}}
\end{picture}
\\
D=8
\vspace{5mm}
\hspace{15mm}
\begin{picture}(280,60)
\thicklines
\put(0,0){\circle{14}}
\put(42,0){\circle*{14}}
\put(84,42){\circle{14}}
\put(42,42){\circle{14}}
\put(0,-15){\makebox(0.4,0.6){\tiny $E_{10}-E_9$}}
\put(42,-15){\makebox(0.4,0.6){\tiny $E_9$}}
\put(8,42){\makebox(0.4,0.6){\tiny $C-E_{10}-E_9$}}
\put(128,42){\makebox(0.4,0.6){\tiny $\tC-E_{10}-E_9$}}
\put(7,0){\line(1,0){28}}
\put(42,7){\line(0,1){28}}
\put(47,5){\line(1,1){32}}
\put(79,37){\line(1,1){10}}\put(79,47){\line(1,-1){10}}
\end{picture}
\\
D=7
\vspace{5mm}
\hspace{15mm}
\begin{picture}(280,60)
\thicklines
\multiput(0,0)(42,0){2}{\circle{14}}
\put(84,42){\circle{14}}
\put(84,42){\circle*{10}}
\put(42,42){\circle{14}}
\put(84,0){\circle*{14}}
\put(0,-15){\makebox(0.4,0.6){\tiny $E_{10}-E_9$}}
\put(42,-15){\makebox(0.4,0.6){\tiny $E_9-E_8$}}
\put(84,-15){\makebox(0.4,0.6){\tiny $E_8$}}
\put(8,42){\makebox(0.4,0.6){\tiny $C-E_{10}-E_9$}}
\put(128,42){\makebox(0.4,0.6){\tiny $\tC-E_{10}-E_9-E_8$}}
\put(7,0){\line(1,0){28}}
\put(49,0){\line(1,0){28}}
\put(42,7){\line(0,1){28}}
\put(84,7){\line(0,1){28}}
\end{picture}
\\
D=6
\vspace{5mm}
\hspace{15mm}
\begin{picture}(280,60)
\thicklines
\multiput(0,0)(42,0){3}{\circle{14}}
\put(126,42){\circle{14}}
\put(42,42){\circle{14}}
\put(126,0){\circle*{14}}
\put(0,-15){\makebox(0.4,0.6){\tiny $E_{10}-E_9$}}
\put(42,-15){\makebox(0.4,0.6){\tiny $E_9-E_8$}}
\put(84,-15){\makebox(0.4,0.6){\tiny $E_8-E_7$}}
\put(126,-15){\makebox(0.4,0.6){\tiny $E_7$}}
\put(77,42){\makebox(0.4,0.6){\tiny $C-E_{10}-E_9$}}
\put(180,42){\makebox(0.4,0.6){\tiny $\tC-E_{10}-E_9-E_8-E_7$}}
\put(7,0){\line(1,0){28}}
\put(49,0){\line(1,0){28}}
\put(91,0){\line(1,0){28}}
\put(42,7){\line(0,1){28}}
\put(126,7){\line(0,1){28}}
\put(121,37){\line(1,1){10}}\put(121,47){\line(1,-1){10}}
\end{picture}
\\
D=5
\vspace{5mm}
\hspace{15mm}
\begin{picture}(280,60)
\thicklines
\multiput(0,0)(42,0){4}{\circle{14}}
\put(168,42){\circle*{14}}
\put(42,42){\circle{14}}
\put(168,0){\circle*{14}}
\put(0,-15){\makebox(0.4,0.6){\tiny $E_{10}-E_9$}}
\put(42,-15){\makebox(0.4,0.6){\tiny $E_9-E_8$}}
\put(84,-15){\makebox(0.4,0.6){\tiny $E_8-E_7$}}
\put(126,-15){\makebox(0.4,0.6){\tiny $E_7-E_6$}}
\put(168,-15){\makebox(0.4,0.6){\tiny $E_6$}}
\put(77,42){\makebox(0.4,0.6){\tiny $C-E_{10}-E_9$}}
\put(230,42){\makebox(0.4,0.6){\tiny $\tC-E_{10}-E_9-E_8-E_7-E_6$}}
\put(7,0){\line(1,0){28}}
\put(49,0){\line(1,0){28}}
\put(91,0){\line(1,0){28}}
\put(133,0){\line(1,0){28}}
\put(42,7){\line(0,1){28}}
\put(168,7){\line(0,1){28}}
\end{picture}
\\
D=4
\vspace{5mm}
\hspace{15mm}
\begin{picture}(280,60)
\thicklines
\multiput(0,0)(42,0){5}{\circle{14}}
\put(210,42){\circle{14}}
\put(42,42){\circle{14}}
\put(210,0){\circle*{14}}
\put(0,-15){\makebox(0.4,0.6){\tiny $E_{10}-E_9$}}
\put(42,-15){\makebox(0.4,0.6){\tiny $E_9-E_8$}}
\put(84,-15){\makebox(0.4,0.6){\tiny $E_8-E_7$}}
\put(126,-15){\makebox(0.4,0.6){\tiny $E_7-E_6$}}
\put(168,-15){\makebox(0.4,0.6){\tiny $E_6-E_5$}}
\put(210,-15){\makebox(0.4,0.6){\tiny $E_5$}}
\put(77,42){\makebox(0.4,0.6){\tiny $C-E_{10}-E_9$}}
\put(282,42){\makebox(0.4,0.6){\tiny $\tC-E_{10}-E_9-E_8-E_7-E_6-E_5$}}
\put(7,0){\line(1,0){28}}
\put(49,0){\line(1,0){28}}
\put(91,0){\line(1,0){28}}
\put(133,0){\line(1,0){28}}
\put(175,0){\line(1,0){28}}
\put(42,7){\line(0,1){28}}
\put(210,7){\line(0,1){28}}
\end{picture}
\\
D=3
\hspace{15mm}
\begin{picture}(280,60)
\thicklines
\multiput(0,0)(42,0){6}{\circle{14}}
\put(210,42){\circle{14}}
\put(42,42){\circle{14}}
\put(252,0){\circle*{14}}
\put(0,-15){\makebox(0.4,0.6){\tiny $E_{10}-E_9$}}
\put(42,-15){\makebox(0.4,0.6){\tiny $E_9-E_8$}}
\put(84,-15){\makebox(0.4,0.6){\tiny $E_8-E_7$}}
\put(126,-15){\makebox(0.4,0.6){\tiny $E_7-E_6$}}
\put(168,-15){\makebox(0.4,0.6){\tiny $E_6-E_5$}}
\put(210,-15){\makebox(0.4,0.6){\tiny $E_5-E_4$}}
\put(252,-15){\makebox(0.4,0.6){\tiny $E_4$}}
\put(77,42){\makebox(0.4,0.6){\tiny $C-E_{10}-E_9$}}
\put(282,42){\makebox(0.4,0.6){\tiny $\tC-E_{10}-E_9-E_8-E_7-E_6-E_5$}}
\put(7,0){\line(1,0){28}}
\put(49,0){\line(1,0){28}}
\put(91,0){\line(1,0){28}}
\put(133,0){\line(1,0){28}}
\put(175,0){\line(1,0){28}}
\put(42,7){\line(0,1){28}}
\put(210,7){\line(0,1){28}}
\put(217,0){\line(1,0){28}}
\end{picture}
\vspace{5mm}

In this section the norms of the imaginary roots and superroots can be
smaller than their values elsewhere in the paper, namely 0 and $-1$.
Let us note also that in $D=3$ the anticanonical class is again 
the sum of the odd simple root and the most positive root of the Lie
algebra $D8$ of U-dualities.

As we said above, the story is the same for Heterotic theories.

\subsection{Fixed subalgebras of automorphisms of IIA/IIB  Borcherds algebras}

In 10 dimensions, the 
Borcherds algebra associated to (truncated) Heterotic supergravity
can be obtained even more easily as the fixed point subalgebra of the following
automorphisms acting on the Borcherds superalgebra of $IIA$ theory:
\bea
e_{\alpha_0} & \rightarrow & + e_{\alpha_0} \nn\\
e_{\alpha_1} & \rightarrow & - e_{\alpha_1} \ .
\eea

This involution preserves the potential $\v$ if $\cA_{(1)} \rightarrow - 
\cA_{(1)}$ and
$\cA_{(2)} \rightarrow  \cA_{(2)}$,
which corresponds to the duality between M-theory compactified on $S^1/
{\mathbb Z}_2$ and Heterotic theory. 
A similar map can be found between $IIB$ supergravity and (truncated)
type $I$ by considering on the $IIB$ Borcherds algebra the automorphism:
\bea
e_{\alpha_0} & \rightarrow & + e_{\alpha_0} \nn\\
e_{\alpha_1} & \rightarrow & - e_{\alpha_1}\ .
\eea
This corresponds to the
orientifold projection. There is a S-dual version of that which maps
to the truncated Heterotic theory:
\bea
e_{\alpha_0} & \rightarrow & - e_{\alpha_0} \nn\\
e_{\alpha_1} & \rightarrow & - e_{\alpha_1} \ .
\eea
We should note that the twisted sectors do not appear and we obtain
the Type $I$ or Heterotic theories without gauge sector. It is
not clear, at present, how to encode the gauge group on the del Pezzo side.

\section{Magic triangle}

\subsection{The triangle}

We have shown that the smooth surface
${\mathbb B}_k$ corresponds to M theory compactified on a $k$ torus.
We have also seen that the dimensional reduction of eleven-dimensional
supergravity to $D=3$ gives rise to a scalar coset theory with an
$E_8$ global symmetry. In \cite{cre}, the oxidation endpoints of
three-dimensional symmetric space scalar theories $G/ KG$ coupled
to gravity have been determined in particular 
for all split subgroups $G$ of split $E_8$.
Split is called maximally noncompact in some of the Physics literature,
for instance $E_5=SO(5,5)$ or $E_4=SL(5,R)$.

By the {\it oxidation endpoint} of a three-dimensional
scalar coset, we mean the bosonic theory in the highest possible
dimension whose toroidal dimensional reduction gives back precisely
the three-dimensional scalar model. It has been shown that the
oxidation endpoint dimension for the subgroup $E_n$ with $2 \leq n
\leq 8$ is usually given by $D_{max}=n+2$ or $n+3$. The oxidation sequence is
presented in the following table and a {\it magic} reflection symmetry
across the diagonal appears.

\bigskip\bigskip
\centerline{
\begin{tabular}{|c|c|c|c|c|c|c|c|c|c|}\hline
&$n=8$&$n=7$&$n=6$&$n=5$&$n=4$&$n=3$&$n=2$&$n=1$&$n=0$\\ \hline
$D=11$ & + & & & & & & & &\\ \hline
$D=10$ &$\R {\it or} A_1$ & +  &&&&&&& \\ \hline  
$D=9$ & $\R\times A_1$ & $\R$ & & &&&&& \\ \hline
$D=8$ & $A_1\times A_2$ & $\R \times A_1$ & $A_1$ & &&&&&\\ \hline
$D=7$ & $E_4$ & $\R \times A_2$ & $\R \times A_1$ & $\R$ & + & & &&\\ \hline
$D=6$ & $E_5$ & $ A_1\times A_3$ & $\R\times A_1^2$
        & $ \R ^2$ & $\R$ & & & &    \\ \hline
$D=5$ & $E_6$ & $A_5$ & $A_2^2$ & $
       \R\times A_1^2$ & $\R \times A_1$ & $A_1$ & & &\\ \hline
$D=4$ & $E_7$ & $D_6$ & $A_5$ & $A_1\times A_3$ &
        $\R \times A_2$ & $\R \times A_1$ & $\R$ &+ &  \\ \hline
$D=3$ & $E_8$ & $E_7$ & $E_6$ & $E_5$ & $E_4$ & $A_1\times A_2$ & $\R\times 
A_1$ & $\R {\it or} A_1$&+\\ \hline
\end{tabular}
            }
\bigskip
\centerline{{\bf Table }: Disintegration (\ie Oxidation) for $E_n$ Cosets}
\bigskip\bigskip 

Each vertical step down corresponds to compactification on a circle.
In all cases, the oxidation endpoint theory includes the metric, a
dilaton and a 3-form potential, and in $D \leq 7$ there are no
additional field potentials. In $D=10$, corresponding to $E_8$, there
is also a 2-form potential and a vector potential. In $D=9$,
corresponding to $E_7$, there is just an additional vector
potential.  In $D=8$,  the $E_6$ column has an additional 0-form potential,
or axion. There are, as we saw, three special cases that arise.  For $E_8$ the
``endpoint'' implied by the generic discussion, namely $D=10$, can be
further oxidized to the bosonic sector of $D=11$ supergravity.  For
$E_7$, the generic discussion leads to an endpoint in $D=9$, but
again a further oxidation is possible, giving in this case a
truncation of type IIB supergravity in $D=10$ to the metric plus the
self-dual 5-form.  One could have predicted a bifurcation above the case
$n=7$ and $D=6$, it is actually not there.
A third special case is $E_4$, for which the
``endpoint'' in $D=6$ can be further oxidized to pure gravity in
$D=7$, after first dualising the 3-form potential to a vector in
$D=6$. 

\subsection{Normal Gorenstein del Pezzo surfaces}

Normal Gorenstein del Pezzo surfaces have been studied by Du Val
\cite{duv}, Demazure
\cite{dem} and
the classification is given in a theorem of Hidaka and Watanabe \cite{hid}:

Denoting by $\pi: \ \tilde{X} \rightarrow X $ a minimal resolution of $X$,
a normal del Pezzo surfaces such that
$H^1(\tilde{X}, {\mathcal O}_{\tilde{X}})=0$
(a natural assumption of connectedness) we have:

(i)  $3 \leq D \leq 11$

(ii)  X is smooth or the singular points of X are rational double points

(iii)  if $D=11$ then $X \simeq {\mathbb P}^2$

(iv) if $D=10$ then either 
$X \simeq {\mathbb P}^1 \times {\mathbb
P}^1$ or $X \simeq {\mathbb B}_1$ or 
$X$ is the cone over a quadric in 
${\mathbb P}^2$ 

(v)  if $ 3 \leq D \leq 9$, then there exists a set $\Sigma$ of points on 
${\mathbb P}^2$ such that the points of $\Sigma$ are in almost general
position, $\vert \Sigma \vert=11-D$ and  $\tilde{X}=V(\Sigma)$. In this
case, the resolution $\pi$ is the contraction of all curves on $\tilde{X}$
with self-intersection $-2$.

The singularities involved here are ``Du Val singularities'' i.e. they are
singular points resulting from the contraction of a set of {\it intersecting}
$(-2)$-curves. Such a singularity is characterized by a Dynkin diagram which 
describes the configuration of the contracted curves and belongs to the ADE
classification \cite{rei2}. Normal del Pezzo surfaces of Picard rank one or
two are almost uniquely determined by their singularity type and are
classified in \cite{mz1,mz2,ye}.

We claim here that the supergravity theories of the magic triangle
correspond to Gorenstein normal del Pezzo surfaces with one $A_p$ singular
point. Precisely, the surfaces of the triangle are exactly the rank
one and rank 
two Gorenstein normal del Pezzo surfaces with one $A_p$ singularity
($p=8-n$) and their generic blow up's.

Let us look first at Picard rank one surfaces with such a
singularity. Five of them have $A_p$ singularity, with $p=0, 1, 4, 7$
and $8$.

For $p=0$, there is no singularity: it is simply $\CP^2$, with
$K^2=9$, and we have already seen that it
corresponds to eleven-dimensional supergravity. We have also seen that
blowing up points in general position gives surfaces corresponding to
toroidal compactifications of this theory.

The $p=1$ case is the cone over a quadric in 
${\mathbb P}^2$, which has indeed $A_1$ singularity and has
$K^2=8$. It corresponds to the truncation of type $IIB$ supergravity in
$D=10$ to the metric plus the
self-dual 5-form . We have $\tilde{X} \simeq  {\mathbb F}_2$ and the
resolution  $\pi$ is given by contracting the minimal section of
$\tilde{X}$. The Hirzebruch surface ${\mathbb F}_2$ is a ${\mathbb P}^1$-bundle
over ${\mathbb P}^1$. Its Picard lattice is generated by its minimal
section $E$, such
that $E^2=-2$, and a fiber $C$. We
know from its bundle structure that $C^2 = 0$ and $E.C=1$. 
The anticanonical divisor can be found using the adjunction
formula: $-K= 2E+4C$ and one may verify that $K^2=8$. The 
Picard lattice of the cone after
contraction of the $(-2)$-curve $E$ is the sublattice orthogonal to $E$ in
$Pic({\mathbb F}_2)$. It is generated by the unique rational
curve $E+2C$, and the anticanonical divisor $-K$ is unchanged.
This rational curve of degree 4  may correspond to a 3-brane
$B3$ coupled to a self-dual 5-form on the field theory side. 
Now, as in the previous section, we
can construct the rank one Borcherds algebra whose truncation should be
a symmetry of the 
corresponding field theory. Then, the whole vertical column 
$n=7$ can be obtained by blowing up generic points on this
quadric cone, which still corresponds to toroidal compactifications of
the $D=10$ theory.

For $p=4$, we start from $\CP^2$ and blow up a point on it. Then take a
point on the exceptional curve of this blow up and blow it up into a
new exceptional curve. Now take again a point on this one, such that
the three considered points sit on a line, and blow it up. After
blowing up a fourth point taken on the last exceptional curve, we have
a surface with four $(-2)$-curves in $A_4$ configuration, which is not
del Pezzo. But contracting this bunch of $(-2)$-curves we get a singular
del Pezzo surface of type $A_4$ with a Picard lattice of rank one. The Picard
lattice is easy to compute: if as usual $H$ is a line in $\CP^2$
and $E_{11}$, $E_{10}$, $E_9$ and $E_8$ are the four exceptional curves.
Now we have also
four $(-2)$-curves representing the classes of 
$E_{11}-E_{10}$, $E_{10}-E_9$, $E_9-E_8$ and
$H-E_{11}-E_{10}-E_9$ and one has $K=-3H+E_{11}+E_{10}+E_9+E_8$
($K^2=5$) as in the regular case. The Picard lattice of the final
surface is the sublattice orthogonal to the four $(-2)$-curves and is
generated by $-K$, which is not changed by this contraction (Du Val 
singularities have crepant resolutions in technical terms). $K^2=5$
tells us that this surface corresponds to a $D=7$ theory, which is pure
gravity.

The next two cases are constructed in the same way. For $p=7$, one
blows up seven points which are taken infinitely close as above, such
that the six divisors $E_{11}-E_{10}$ to $E_6-E_5$ are
$(-2)$-curves; the first six points must also lie on a conic $2H$, so
that $2H-E_{11}-E_{10}-E_9-E_8-E_7-E_6$ is also a $(-2)$-curve. All these
$(-2)$-curves form an $A_7$ diagram, and after contraction of them we
have a del Pezzo surface with $A_7$ singularity and a Picard lattice
generated by $-K=3H-E_{11}-E_{10}-E_9-E_8-E_7-E_6-E_5$. As $K^2=2$, it
corresponds to a $D=4$ theory, which is simply four-dimensional gravity.

For $p=8$, the story is the same with eight infinitely close points
blown up such that $E_{11}-E_{10}$ to $E_5-E_4$ are (-2)-curves and
$3H-2E_{11}-E_{10}-E_9-E_8-E_7-E_6-E_5-E_4$ is too. This last condition
means that there is conic $3H$ which passes through all the considered
points and has the first one as a double point. Contracting all these
(-2)-curves, one gets this time an $A_8$-singular del Pezzo surface
with Picard lattice generated by $-K$, with $D=K^2+2=3$.

We may note the general formula $K^2=n+1$ in the minimal  rank one case. 
Let us now look at Picard rank two del Pezzo surfaces with $A_p$
singularity. Some of them are simply obtained by blowing up a generic
point on one of the Picard rank one surfaces just described. There is
also $\CP^1 \times \CP^1$ which corresponds to $IIB$ supergravity as we
have seen. The other ones are obtained by blowing up $p+1$
infinitely close points on $\CP^2$, each one lying on the exceptional
curve of the preceding blow-up, so that $E_i-E_{i-1}$'s ($11-p \leq i
\leq 11$) form an $A_p$ sequence of $(-2)$-curves which are then contracted.
One gets thus del Pezzo surfaces with $A_p$ singularity, Picard rank
two and $D=K^2+2=10-p=n+2$.

All other surfaces of the magic triangle are obtained by blowing up
those of Picard rank one and two, $A_p$-singular, del Pezzo surfaces, which 
are still related to toroidal compactifications of the corresponding
field theories. One might expect that in a given column blowing up
points on del Pezzo's of Picard rank one and two generates two sequences
of surfaces. But this is not the case: we already know that $IIA$ (${\mathbb
B}_1$) and $IIB$ ($\CP^1 \times \CP^1$) give the same surface when
compactified on tori. Similarly, blowing up a generic point on the
quadric cone for $n=7$ and on the 
$A_4$ Picard rank one del Pezzo gives respectively the same surfaces
as the Picard 
rank two surfaces described in the previous paragraph with $p=1$ and $p=4$. 
Actually we have general relations:
$$ D-2=K^2=n+1-k-r=9-r-(k+p) $$
where $r$ is the rank of the Picard group after blowing down all $-1$-curves.

There are three more cases with two different del
Pezzo's beyond those corresponding to U-duality groups $A_1$ and ${\mathbb R}$:
$n=8$, $D=10$ and $n=1$, $D=3$. One can indeed get $A_2$ instead of
${\mathbb R} \times A_1$ in $n=7$, $D=8$ and $n=3$, $D=4$, and one can
have $A_1^2$ in addition to the ${\mathbb R}^2$ case in $n=5$, $D=6$.

\subsection{Symmetric triangles}

The symmetry of the triangle with respect to reflection across the
diagonal relates two  field theories with identical U-duality groups. On
the geometric side, this is a symmetry between the set of
$(-2)$-divisors of degree 0, and therefore between the Weyl groups
they generate. Actually, we can prove that the sublattices orthogonal
to $-K$ in the Picard lattices of any pair of symmetric surfaces of the
triangle are identical.

This can be explained by the fact that on a del Pezzo surface of
degree $K^2 = 1$, the orthogonal to $n$ mutually orthogonal $(-1)$ rational
divisors and the orthogonal to an $A_n$ chain of $(-2)$ rational divisors
are the same, when we restrict attention to divisors orthogonal to $-K$
in the Picard lattice. Therefore contracting $n$ orthogonal
$(-1)$-curves and taking the orthogonal to an $A_k$ chain of $(-2)$-divisors
gives the same degree 0 part of the Picard lattice as what one gets with
$n$ and $k$ exchanged. Consequently the $K^\perp$'s are the same for the
del Pezzo of degree $(1+n)$ with an $A_k$  singularity and the del Pezzo
of degree $(1+k)$ with singularity $A_n$.\footnote{A. Keurentjes informed us 
that he could trace the symmetry of the
triangle to the possibility of extracting a linear subgroup of dualities
from each  side of the triangle, we independently proved the symmetry from 
algebraic geometry and related together three different mechanisms, the 
singular  blow-downs,  the regular blow-downs and the older observation of 
linear subgroups from dimensional reduction.}

This argument applies to any del Pezzo of degree one and not only to
the smooth one, so one can construct a similar triangle starting from
any of them, for instance there is a type I magic triangle, we shall return to
these in a subsequent publication.

\subsection{Dynkin diagrams}

We give here Borcherds superalgebras attached to the $A_{p}$-singular normal
del Pezzo's from which (the bosonic part of) field theories  can be retrieved
($p=8-n$).
In order to calculate the lengths of simple roots, one must know that
$B_i^2=i-1$ (these divisors correspond to $i$-branes).
When the only root of the Dynkin diagram is $-K$, its multiplicity
must be set to zero to be consistent with Hodge duality.
\vfill
\eject
\subsubsection{$n=7$}

\noindent

D=10 
\vspace{5mm}
\hspace{13mm}
\begin{picture}(120,18)
\thicklines
\put(42,0){\circle{14}}
\put(37,-5){\line(1,1){10}}\put(37,5){\line(1,-1){10}}
\put(42,-15){\makebox(0.4,0.6){\tiny $-K/2=B_3$}}
\end{picture}
\vspace{5mm}
  
D=9
\vspace{5mm}
\hspace{15mm}
\begin{picture}(250,18)
\thicklines
\put(0,0){\circle*{14}}
\put(42,0){\circle{14}}
\put(42,0){\circle*{10}}
\put(0,-15){\makebox(0.4,0.6){\tiny $B_0$}}
\put(42,-15){\makebox(0.4,0.6){\tiny $B_2$}}
\put(7,0){\line(1,0){28}}
\end{picture}
\vspace{5mm}
 
D=8 
\vspace{5mm}
\hspace{15mm}
\begin{picture}(250,60)
\thicklines
\put(0,0){\circle{14}}
\put(42,42){\circle{14}}
\put(42,0){\circle*{14}}
\put(0,-15){\makebox(0.4,0.6){\tiny $B_0-E_9$}}
\put(42,-15){\makebox(0.4,0.6){\tiny $E_9$}}
\put(70,42){\makebox(0.4,0.6){\tiny $B_2-E_9$}}
\put(7,0){\line(1,0){28}}
\put(42,7){\line(0,1){28}}
\put(37,37){\line(1,1){10}}\put(37,47){\line(1,-1){10}}
\end{picture}
\vspace{5mm}
 
D=7 
\vspace{5mm}
\hspace{15mm}
\begin{picture}(250,60)
\thicklines
\multiput(0,0)(42,0){2}{\circle{14}}
\put(84,42){\circle*{14}}
\put(84,0){\circle*{14}}
\put(0,-15){\makebox(0.4,0.6){\tiny $B_0-E_9$}}
\put(42,-15){\makebox(0.4,0.6){\tiny $E_9-E_8$}}
\put(84,-15){\makebox(0.4,0.6){\tiny $E_8$}}
\put(120,42){\makebox(0.4,0.6){\tiny $B_2-E_9-E_8$}}
\put(7,0){\line(1,0){28}}
\put(49,0){\line(1,0){28}}
\put(84,7){\line(0,1){28}}
\end{picture}
\vspace{5mm}

D=6 
\vspace{5mm}
\hspace{15mm}
\begin{picture}(250,60)
\thicklines
\multiput(0,0)(42,0){3}{\circle{14}}
\put(126,42){\circle{14}}
\put(126,0){\circle*{14}}
\put(0,-15){\makebox(0.4,0.6){\tiny $B_0-E_9$}}
\put(42,-15){\makebox(0.4,0.6){\tiny $E_9-E_8$}}
\put(84,-15){\makebox(0.4,0.6){\tiny $E_8-E_7$}}
\put(126,-15){\makebox(0.4,0.6){\tiny $E_7$}}
\put(170,42){\makebox(0.4,0.6){\tiny $B_2-E_9-E_8-E_7$}}
\put(7,0){\line(1,0){28}}
\put(49,0){\line(1,0){28}}
\put(91,0){\line(1,0){28}}
\put(126,7){\line(0,1){28}}
\end{picture}
\vspace{5mm}

D=5 
\vspace{5mm}
\hspace{15mm}
\begin{picture}(250,60)
\thicklines
\multiput(0,0)(42,0){4}{\circle{14}}
\put(126,42){\circle{14}}
\put(168,0){\circle*{14}}
\put(0,-15){\makebox(0.4,0.6){\tiny $B_0-E_9$}}
\put(42,-15){\makebox(0.4,0.6){\tiny $E_9-E_8$}}
\put(84,-15){\makebox(0.4,0.6){\tiny $E_8-E_7$}}
\put(126,-15){\makebox(0.4,0.6){\tiny $E_7-E_6$}}
\put(168,-15){\makebox(0.4,0.6){\tiny $E_6$}}
\put(170,42){\makebox(0.4,0.6){\tiny $B_2-E_9-E_8-E_7$}}
\put(7,0){\line(1,0){28}}
\put(49,0){\line(1,0){28}}
\put(91,0){\line(1,0){28}}
\put(133,0){\line(1,0){28}}
\put(126,7){\line(0,1){28}}
\end{picture}
\vspace{5mm}
 
D=4 
\vspace{5mm}
\hspace{15mm}
\begin{picture}(250,60)
\thicklines
\multiput(0,0)(42,0){5}{\circle{14}}
\put(126,42){\circle{14}}
\put(210,0){\circle*{14}}
\put(0,-15){\makebox(0.4,0.6){\tiny $B_0-E_9$}}
\put(42,-15){\makebox(0.4,0.6){\tiny $E_9-E_8$}}
\put(84,-15){\makebox(0.4,0.6){\tiny $E_8-E_7$}}
\put(126,-15){\makebox(0.4,0.6){\tiny $E_7-E_6$}}
\put(168,-15){\makebox(0.4,0.6){\tiny $E_6-E_5$}}
\put(210,-15){\makebox(0.4,0.6){\tiny $E_5$}}
\put(170,42){\makebox(0.4,0.6){\tiny $B_2-E_9-E_8-E_7$}}
\put(7,0){\line(1,0){28}}
\put(49,0){\line(1,0){28}}
\put(91,0){\line(1,0){28}}
\put(133,0){\line(1,0){28}}
\put(175,0){\line(1,0){28}}
\put(126,7){\line(0,1){28}}
\end{picture}
\vspace{5mm}
 
D=3 
\vspace{5mm}
\hspace{15mm}
\begin{picture}(250,60)
\thicklines
\multiput(0,0)(42,0){6}{\circle{14}}
\put(126,42){\circle{14}}
\put(252,0){\circle*{14}}
\put(0,-15){\makebox(0.4,0.6){\tiny $B_0-E_9$}}
\put(42,-15){\makebox(0.4,0.6){\tiny $E_9-E_8$}}
\put(84,-15){\makebox(0.4,0.6){\tiny $E_8-E_7$}}
\put(126,-15){\makebox(0.4,0.6){\tiny $E_7-E_6$}}
\put(168,-15){\makebox(0.4,0.6){\tiny $E_6-E_5$}}
\put(210,-15){\makebox(0.4,0.6){\tiny $E_5-E_4$}}
\put(252,-15){\makebox(0.4,0.6){\tiny $E_4$}}
\put(170,42){\makebox(0.4,0.6){\tiny $B_2-E_9-E_8-E_7$}}
\put(7,0){\line(1,0){28}}
\put(49,0){\line(1,0){28}}
\put(91,0){\line(1,0){28}}
\put(133,0){\line(1,0){28}}
\put(175,0){\line(1,0){28}}
\put(126,7){\line(0,1){28}}
\put(217,0){\line(1,0){28}}
\end{picture}
\vspace{5mm}

\subsubsection{$n=6$}
\noindent

D=8
\vspace{5mm}
\hspace{15mm}
\begin{picture}(200,18)
\thicklines
\put(42,0){\circle{14}}
\put(84,0){\circle{14}}
\put(84,0){\circle*{10}}
\put(42,-15){\makebox(0.4,0.6){\tiny $B_{-1}$}}
\put(84,-15){\makebox(0.4,0.6){\tiny $B_2$}}
\put(49,0){\line(1,0){28}}
\end{picture}
\vspace{5mm}
 
D=7 
\vspace{5mm}
\hspace{15mm}
\begin{picture}(200,60)
\thicklines
\put(84,42){\circle{14}}
\put(42,42){\circle{14}}
\put(84,0){\circle*{14}}
\put(84,-15){\makebox(0.4,0.6){\tiny $E_8$}}
\put(110,42){\makebox(0.4,0.6){\tiny $B_2-E_8$}}
\put(23,42){\makebox(0.4,0.6){\tiny $B_{-1}$}}
\put(49,42){\line(1,0){28}}
\put(84,7){\line(0,1){28}}
\put(79,37){\line(1,1){10}}\put(79,47){\line(1,-1){10}}
\end{picture}
\vspace{5mm}
 
D=6 
\vspace{5mm}
\hspace{15mm}
\begin{picture}(200,60)
\thicklines
\put(42,0){\circle{14}}
\put(84,42){\circle*{14}}
\put(42,42){\circle{14}}
\put(84,0){\circle*{14}}
\put(42,-15){\makebox(0.4,0.6){\tiny $E_8-E_7$}}
\put(84,-15){\makebox(0.4,0.6){\tiny $E_7$}}
\put(117,42){\makebox(0.4,0.6){\tiny $B_2-E_8-E_7$}}
\put(23,42){\makebox(0.4,0.6){\tiny $B_{-1}$}}
\put(49,0){\line(1,0){28}}
\put(49,42){\line(1,0){28}}
\put(84,7){\line(0,1){28}}
\end{picture}
\vspace{5mm}
 
D=5 
\vspace{5mm}
\hspace{15mm}
\begin{picture}(200,60)
\thicklines
\multiput(0,0)(42,0){2}{\circle{14}}
\put(84,42){\circle{14}}
\put(42,42){\circle{14}}
\put(84,0){\circle*{14}}
\put(0,-15){\makebox(0.4,0.6){\tiny $E_8-E_7$}}
\put(42,-15){\makebox(0.4,0.6){\tiny $E_7-E_6$}}
\put(84,-15){\makebox(0.4,0.6){\tiny $E_6$}}
\put(127,42){\makebox(0.4,0.6){\tiny $B_2-E_8-E_7-E_6$}}
\put(23,42){\makebox(0.4,0.6){\tiny $B_{-1}$}}
\put(7,0){\line(1,0){28}}
\put(49,0){\line(1,0){28}}
\put(49,42){\line(1,0){28}}
\put(84,7){\line(0,1){28}}
\end{picture}
\vspace{5mm}
 
D=4 
\vspace{5mm}
\hspace{15mm}
\begin{picture}(200,60)
\thicklines
\multiput(0,0)(42,0){3}{\circle{14}}
\put(84,42){\circle{14}}
\put(42,42){\circle{14}}
\put(126,0){\circle*{14}}
\put(0,-15){\makebox(0.4,0.6){\tiny $E_8-E_7$}}
\put(42,-15){\makebox(0.4,0.6){\tiny $E_7-E_6$}}
\put(84,-15){\makebox(0.4,0.6){\tiny $E_6-E_5$}}
\put(126,-15){\makebox(0.4,0.6){\tiny $E_5$}}
\put(127,42){\makebox(0.4,0.6){\tiny $B_2-E_8-E_7-E_6$}}
\put(23,42){\makebox(0.4,0.6){\tiny $B_{-1}$}}
\put(7,0){\line(1,0){28}}
\put(49,0){\line(1,0){28}}
\put(91,0){\line(1,0){28}}
\put(49,42){\line(1,0){28}}
\put(84,7){\line(0,1){28}}
\end{picture}
\vspace{5mm}
 
D=3 
\vspace{5mm}
\hspace{15mm}
\begin{picture}(200,60)
\thicklines
\multiput(0,0)(42,0){4}{\circle{14}}
\put(84,42){\circle{14}}
\put(42,42){\circle{14}}
\put(168,0){\circle*{14}}
\put(0,-15){\makebox(0.4,0.6){\tiny $E_8-E_7$}}
\put(42,-15){\makebox(0.4,0.6){\tiny $E_7-E_6$}}
\put(84,-15){\makebox(0.4,0.6){\tiny $E_6-E_5$}}
\put(126,-15){\makebox(0.4,0.6){\tiny $E_5-E_4$}}
\put(168,-15){\makebox(0.4,0.6){\tiny $E_4$}}
\put(127,42){\makebox(0.4,0.6){\tiny $B_2-E_8-E_7-E_6$}}
\put(23,42){\makebox(0.4,0.6){\tiny $B_{-1}$}}
\put(7,0){\line(1,0){28}}
\put(49,0){\line(1,0){28}}
\put(91,0){\line(1,0){28}}
\put(133,0){\line(1,0){28}}
\put(49,42){\line(1,0){28}}
\put(84,7){\line(0,1){28}}
\end{picture}
\vspace{5mm}

\subsubsection{$n=5$}
\noindent

D=7 
\vspace{5mm}
\hspace{15mm}
\begin{picture}(200,18)
\thicklines
\put(42,0){\circle{14}}
\put(84,0){\circle{14}}
\put(84,0){\circle*{10}}
\put(42,-15){\makebox(0.4,0.6){\tiny $B_1$}}
\put(84,-15){\makebox(0.4,0.6){\tiny $B_2$}}
\put(49,0){\line(1,0){28}}
\put(37,-5){\line(1,1){10}}\put(37,5){\line(1,-1){10}}
\end{picture}
\vspace{5mm}

D=6 
\vspace{5mm}
\hspace{15mm}
\begin{picture}(150,60)
\thicklines
\put(84,42){\circle{14}}
\put(42,42){\circle*{14}}
\put(42,0){\circle*{14}}
\put(42,-15){\makebox(0.4,0.6){\tiny $E_7$}}
\put(110,42){\makebox(0.4,0.6){\tiny $B_2-E_7$}}
\put(17,42){\makebox(0.4,0.6){\tiny $B_1-E_7$}}
\put(42,7){\line(0,1){28}}
\put(47,5){\line(1,1){32}}
\put(79,37){\line(1,1){10}}\put(79,47){\line(1,-1){10}}
\end{picture}
\vspace{5mm}

D=5 
\vspace{5mm}
\hspace{15mm}
\begin{picture}(150,60)
\thicklines
\put(0,0){\circle{14}}
\put(84,42){\circle*{14}}
\put(42,42){\circle{14}}
\put(42,0){\circle*{14}}
\put(0,-15){\makebox(0.4,0.6){\tiny $E_7-E_6$}}
\put(42,-15){\makebox(0.4,0.6){\tiny $E_6$}}
\put(120,42){\makebox(0.4,0.6){\tiny $B_2-E_7-E_6$}}
\put(7,42){\makebox(0.4,0.6){\tiny $B_1-E_7-E_6$}}
\put(7,0){\line(1,0){28}}
\put(42,7){\line(0,1){28}}
\put(47,5){\line(1,1){32}}
\end{picture}
\vspace{5mm}

D=4 
\vspace{5mm}
\hspace{15mm}
\begin{picture}(150,60)
\thicklines
\multiput(0,0)(42,0){2}{\circle{14}}
\put(84,42){\circle{14}}
\put(42,42){\circle{14}}
\put(84,0){\circle*{14}}
\put(0,-15){\makebox(0.4,0.6){\tiny $E_7-E_6$}}
\put(42,-15){\makebox(0.4,0.6){\tiny $E_6-E_5$}}
\put(84,-15){\makebox(0.4,0.6){\tiny $E_5$}}
\put(127,42){\makebox(0.4,0.6){\tiny $B_2-E_7-E_6-E_5$}}
\put(7,42){\makebox(0.4,0.6){\tiny $B_1-E_7-E_6$}}
\put(7,0){\line(1,0){28}}
\put(49,0){\line(1,0){28}}
\put(42,7){\line(0,1){28}}
\put(84,7){\line(0,1){28}}
\end{picture}
\vspace{5mm}

D=3 
\vspace{5mm}
\hspace{15mm}
\begin{picture}(150,60)
\thicklines
\multiput(0,0)(42,0){3}{\circle{14}}
\put(84,42){\circle{14}}
\put(42,42){\circle{14}}
\put(126,0){\circle*{14}}
\put(0,-15){\makebox(0.4,0.6){\tiny $E_7-E_6$}}
\put(42,-15){\makebox(0.4,0.6){\tiny $E_6-E_5$}}
\put(84,-15){\makebox(0.4,0.6){\tiny $E_5-E_4$}}
\put(126,-15){\makebox(0.4,0.6){\tiny $E_4$}}
\put(127,42){\makebox(0.4,0.6){\tiny $B_2-E_7-E_6-E_5$}}
\put(7,42){\makebox(0.4,0.6){\tiny $B_1-E_7-E_6$}}
\put(7,0){\line(1,0){28}}
\put(49,0){\line(1,0){28}}
\put(91,0){\line(1,0){28}}
\put(42,7){\line(0,1){28}}
\put(84,7){\line(0,1){28}}
\end{picture}
\vspace{5mm}

\subsubsection{$n=4$}
\noindent

D=7 
\vspace{5mm}
\hspace{15mm}
\begin{picture}(150,18)
\thicklines
\put(84,0){\circle{14}}
\put(84,0){\circle*{10}}
\put(84,-15){\makebox(0.4,0.6){\tiny $-K$}}
\end{picture}
\vspace{5mm}
 
D=6 
\vspace{5mm}
\hspace{15mm}
\begin{picture}(150,18)
\thicklines
\put(42,0){\circle*{14}}
\put(84,0){\circle{14}}
\put(84,0){\circle*{10}}
\put(42,-15){\makebox(0.4,0.6){\tiny $B_0$}}
\put(84,-15){\makebox(0.4,0.6){\tiny $B_2$}}
\put(48,3){\line(1,0){30}}
\put(48,-3){\line(1,0){30}}
\end{picture}
\vspace{5mm}
 
D=5 
\vspace{5mm}
\hspace{15mm}
\begin{picture}(150,60)
\thicklines
\put(84,42){\circle{14}}
\put(42,42){\circle{14}}
\put(42,0){\circle*{14}}
\put(42,-15){\makebox(0.4,0.6){\tiny $E_6$}}
\put(108,42){\makebox(0.4,0.6){\tiny $B_2-E_6$}}
\put(17,42){\makebox(0.4,0.6){\tiny $B_0-E_6$}}
\put(42,7){\line(0,1){28}}
\put(47,5){\line(1,1){32}}
\put(49,42){\line(1,0){28}}
\put(79,37){\line(1,1){10}}\put(79,47){\line(1,-1){10}}
\end{picture}
\vspace{5mm}
 
D=4 
\vspace{5mm}
\hspace{15mm}
\begin{picture}(150,60)
\thicklines
\put(84,42){\circle*{14}}
\put(42,42){\circle{14}}
\put(42,0){\circle{14}}
\put(84,0){\circle*{14}}
\put(42,-15){\makebox(0.4,0.6){\tiny $E_6-E_5$}}
\put(84,-15){\makebox(0.4,0.6){\tiny $E_5$}}
\put(117,42){\makebox(0.4,0.6){\tiny $B_2-E_6-E_5$}}
\put(17,42){\makebox(0.4,0.6){\tiny $B_0-E_6$}}
\put(42,7){\line(0,1){28}}
\put(84,7){\line(0,1){28}}
\put(49,0){\line(1,0){28}}
\put(49,42){\line(1,0){28}}
\end{picture}
\vspace{5mm}
 
D=3 
\vspace{5mm}
\hspace{15mm}
\begin{picture}(150,60)
\thicklines
\put(84,42){\circle{14}}
\put(42,42){\circle{14}}
\multiput(42,0)(42,0){2}{\circle{14}}
\put(126,0){\circle*{14}}
\put(42,-15){\makebox(0.4,0.6){\tiny $E_6-E_5$}}
\put(84,-15){\makebox(0.4,0.6){\tiny $E_5-E_4$}}
\put(126,-15){\makebox(0.4,0.6){\tiny $E_4$}}
\put(125,42){\makebox(0.4,0.6){\tiny $B_2-E_6-E_5-E_4$}}
\put(17,42){\makebox(0.4,0.6){\tiny $B_0-E_6$}}
\put(42,7){\line(0,1){28}}
\put(91,0){\line(1,0){28}}
\put(49,0){\line(1,0){28}}
\put(49,42){\line(1,0){28}}
\put(89,37){\line(1,-1){32}}
\end{picture}
\vspace{5mm}

\subsubsection{$n=3$}
\noindent

D=5 
\vspace{5mm}
\hspace{15mm}
\begin{picture}(120,18)
\thicklines
\put(42,0){\circle{14}}
\put(84,0){\circle{14}}
\put(84,0){\circle*{10}}
\put(42,-15){\makebox(0.4,0.6){\tiny $B_{-1}$}}
\put(84,-15){\makebox(0.4,0.6){\tiny $B_2$}}
\put(48,3){\line(1,0){30}}\put(48,-3){\line(1,0){30}}
\end{picture}
\vspace{5mm}
 
D=4 
\vspace{5mm}
\hspace{15mm}
\begin{picture}(120,60)
\thicklines
\put(84,42){\circle{14}}
\put(42,42){\circle{14}}
\put(84,0){\circle*{14}}
\put(84,-15){\makebox(0.4,0.6){\tiny $B_2+B_{-1}-2E_5$}}
\put(108,42){\makebox(0.4,0.6){\tiny $B_2-E_5$}}
\put(23,42){\makebox(0.4,0.6){\tiny $B_{-1}$}}
\put(84,7){\line(0,1){28}}
\put(48,45){\line(1,0){30}}\put(48,39){\line(1,0){30}}
\put(79,37){\line(1,1){10}}\put(79,47){\line(1,-1){10}}
\end{picture}
\vspace{5mm}
 
D=3 
\vspace{5mm}
\hspace{15mm}
\begin{picture}(220,18)
\thicklines
\put(84,0){\circle*{14}}
\put(42,0){\circle{14}}
\put(166,0){\circle{14}}
\put(208,0){\circle{14}}
\put(218,-15){\makebox(0.4,0.6){\tiny $E_5-E_4$}}
\put(156,-15){\makebox(0.4,0.6){\tiny $B_2+B_{-1}-2E_5-E_4$}}
\put(80,-15){\makebox(0.4,0.6){\tiny $B_2-E_5-E_4$}}
\put(42,-15){\makebox(0.4,0.6){\tiny $B_{-1}$}}
\put(48,-3){\line(1,0){30}}\put(48,3){\line(1,0){30}}
\put(173,0){\line(1,0){28}}
\end{picture}
\vspace{5mm}

\subsubsection{$n=2$}
\noindent

D=4 
\vspace{5mm}
\hspace{15mm}
\begin{picture}(120,18)
\thicklines
\put(42,0){\circle{14}}
\put(84,0){\circle{14}}
\put(84,0){\circle*{10}}
\put(84,-15){\makebox(0.4,0.6){\tiny $B_2$}}
\put(42,-15){\makebox(0.4,0.6){\tiny $-K$}}
\put(48,4){\line(1,0){30}}\put(48,-4){\line(1,0){30}}\put(49,0){\line(1,0){28}}
\put(37,-5){\line(1,1){10}}\put(37,5){\line(1,-1){10}}
\end{picture}
\vspace{5mm}
 
D=3 
\vspace{5mm}
\hspace{15mm}
\begin{picture}(120,60)
\thicklines
\put(84,42){\circle{14}}
\put(42,0){\circle*{14}}
\put(42,42){\circle{14}}
\put(42,-15){\makebox(0.4,0.6){\tiny $E_4$}}
\put(108,42){\makebox(0.4,0.6){\tiny $B_2-E_4$}}
\put(17,42){\makebox(0.4,0.6){\tiny $-K-E_4$}}
\put(39,6){\line(0,1){30}}\put(45,6){\line(0,1){30}}
\put(49,42){\line(1,0){28}}
\put(47,5){\line(1,1){32}}
\put(37,37){\line(1,1){10}}\put(37,47){\line(1,-1){10}}
\end{picture}
\vspace{5mm}

\subsubsection{$n=1$}
\noindent

D=4
\vspace{5mm}
\hspace{15mm}
\begin{picture}(120,18)
\thicklines
\put(42,0){\circle{14}}
\put(37,-5){\line(1,1){10}}\put(37,5){\line(1,-1){10}}
\put(42,-15){\makebox(0.4,0.6){\tiny $-K$}}
\end{picture}
\vspace{5mm}
 
D=3 
\vspace{5mm}
\hspace{15mm}
\begin{picture}(120,18)
\thicklines
\put(42,0){\circle{14}}
\put(84,0){\circle*{14}}
\put(42,-15){\makebox(0.4,0.6){\tiny $-K-E_4$}}
\put(84,-15){\makebox(0.4,0.6){\tiny $E_4$}}
\put(48,3){\line(1,0){30}}
\put(48,-3){\line(1,0){30}}
\end{picture}
\vspace{5mm}
\hspace{5mm} or \hspace{5mm}
\vspace{5mm}
\begin{picture}(120,18)
\thicklines
\put(42,0){\circle{14}}
\put(42,0){\circle*{10}}
\put(84,0){\circle{14}}
\put(84,0){\circle*{10}}
\put(84,-15){\makebox(0.4,0.6){\tiny $B_2$}}
\put(42,-15){\makebox(0.4,0.6){\tiny $-K$}}
\put(48,4){\line(1,0){30}}\put(48,-4){\line(1,0){30}}\put(49,0){\line(1,0){28}}
\end{picture}
\vspace{5mm}

\subsubsection{$n=0$}
\noindent

D=3
\vspace{5mm}
\hspace{15mm}
\begin{picture}(60,18)
\thicklines
\put(42,0){\circle{14}}
\put(42,0){\circle*{10}}
\put(42,-15){\makebox(0.4,0.6){\tiny $-K$}}
\end{picture}
\vspace{5mm}

Note that in all the diagrams of rank one where the simple root is $-K$ one is 
actually considering pure gravity. 
Again there are some purely bosonic examples of Borcherds algebras.

\section{Conclusion}

In \cite{hm,hm1,dij,kaw}, it was noticed that 
modular integrals associated with one-loop calculations
of compactified heterotic string theory can be interpreted
 as the Weyl denominator formula of a Borcherds (super)-algebra. We recall 
that the Weyl group is by definition generated by symmetries with respect to 
real roots and  this denominator \cite{ray} can be written in two 
equivalent ways as: 
\be
\frac{ \prod_{\alpha \ {\mathrm even}} (1-\alpha)^{{\mathrm mult}(\alpha)}}
{ \prod_{\alpha \ {\mathrm odd}} (1+\alpha)^{{\mathrm
mult}(\alpha)}}=e^{\rho}\sum_{w \in W}{\mathrm det}(w) \, w(e^{-\rho}
\sum_{\mu}\epsilon(\mu)e^{\mu}) 
\ee
where the products on the left handside run over positive roots $\alpha$,
respectively bosonic and fermionic, with multiplicities ${\mathrm
mult}(\alpha)$,
$\rho$ is the Weyl vector, defined by
$(\rho,\alpha_i)=\frac{1}{2}(\alpha_i,\alpha_i)$, $W$ is the Weyl
group, the sum on the right handside runs over $\mu$ itself sum of mutually 
orthogonal imaginary simple roots 
$$\mu=\sum_j \alpha_{i_j {\mathrm even}} + 
\sum_k l_{i_k} \alpha_{i_k {\mathrm odd}}$$ 
for 
positive integers $l_{i_k}$ multiplying the odd imaginary roots restricted
by the condition that
$l_{i_k} \geq 2$ is allowed only if $\alpha_{i_k {\mathrm odd}}$ is isotropic 
i.e. 
$(\alpha_{i_k {\mathrm odd}},\alpha_{i_k {\mathrm odd}})=0$ and $\epsilon(\mu)=(-1)^{{\mathrm ht}(\mu)}$.

One can implement the defining relations of Borcherds algebras most 
efficiently by using recursively the above formula. 
We only need a finite number of checks and this is a finite process. 
The singularities of the denominator function, studied by \cite{bor1}, have been
interpreted in \cite{hm} as the enhanced symmetry points in the even
self-dual Narain lattice. We plan to return to this problem. 

In order to summarize 
we may say that we have considerably enlarged the finite dimensional duality 
superalgebras that themselves were a generalization of the U-duality symmetries.
We may now return to the many open questions about the structure of Einstein, 
supergravity and superstring equations with hope of even more stunning beauty.  
Clearly the real algebraic geometry will be important, the fermionic extensions
will be closely related to it. We may note that the positive degree obstacle
from differential forms has been removed so we may go beyond Borel subgroups 
again and deal with the spacetime fermions. The symmetry of the Magic triangle 
fits in a considerable body of dualities. 
Of course the twistor like construction of 
spacetime as 
a derived object is an open problem but one can see where to look for it 
already.

\section*{Acknowledgments}

    We are grateful to D. Auroux, A. Beauville, J.B. Bost, P. Cartier, 
I. Dolgachev, F. Loeser, D. Madore, W. Nahm, 
D. Naie, V. Nikulin, S. Rozensztajn,
P. Sorba, B. Teissier and D. Zvonkine for useful explanations. B.J. would like 
to thank
Tamas Hauer, Conan Leung and Valery Gritsenko for early discussions on this 
magics in algebraic geometry.

\end{document}